\documentclass[fleqn,usenatbib]{mnras}

\usepackage{newtxtext,newtxmath,amsmath}

\usepackage[T1]{fontenc}

\DeclareRobustCommand{\VAN}[3]{#2}
\let\VANthebibliography\thebibliography
\def\thebibliography{\DeclareRobustCommand{\VAN}[3]{##3}\VANthebibliography}

\usepackage{graphicx}	
\usepackage[dvipsnames]{xcolor}
\usepackage{ulem}

\newcommand{\msun}{\textnormal{M}_\odot}
\newcommand{\rsun}{\textnormal{R}_\odot}

\title[SNe in AGN discs]{Supernova explosions in active galactic nuclear discs}

\author[Grishin et al.]{
Evgeni Grishin$^{1,3,4}$\thanks{E-mail: eugeneg@campus.technion.ac.il (EG)},
Alexey Bobrick$^{2}$, Ryosuke Hirai$^{3,4}$, Ilya Mandel$^{3,4,5}$, Hagai B. Perets $^{1}$ 
\\
$^{1}$Physics Department, Technion - Israel Institute of Technology, 3200003, Haifa, Israel\\
$^{2}$ Lund Observatory, Department of Astronomy and Theoretical Physics, Lund, Sweden \\
$^{3}$ Monash Centre for Astrophysics, School of Physics and Astronomy, Monash University, Clayton, Victoria 3800, Australia\\
$^{4}$ OzGrav: Australian Research Council Centre of Excellence for Gravitational Wave Discovery, Clayton, VIC 3800, Australia\\
$^{5}$ Institute of Gravitational Wave Astronomy and School of Physics and Astronomy, University of Birmingham, Birmingham, B15 2TT, United Kingdom\\
}

\date{Accepted XXX. Received YYY; in original form ZZZ}

\pubyear{2021}

\begin{document}
\defcitealias{yalinewich_matzner2019}{YM19}
\defcitealias{sg03}{SG03}
\defcitealias{MM99}{MM99}

\label{firstpage}
\pagerange{\pageref{firstpage}--\pageref{lastpage}}
\maketitle

\begin{abstract}
Active galactic nuclei (AGN) are prominent environments for stellar capture, growth and formation. These environments may catalyze stellar mergers and explosive transients, such as thermonuclear and core-collapse supernovae (SNe).  
SN explosions in AGN discs generate strong shocks, leading to unique observable signatures.
We develop an analytical model which follows the evolution of the shock propagating in the disc until it eventually breaks out. We derive the peak luminosity, bolometric lightcurve, and breakout time.  The peak luminosities may exceed $10^{45}$ erg s$^{-1}$ and last from hours to days. The brightest explosions occur in regions of reduced density; either off-plane, or in discs around low-mass central black holes ($\sim 10^6\ \msun$), or in starved subluminous AGNs. Explosions in the latter two sites are easier to observe due to a reduced AGN background luminosity. We perform suites of 1D Lagrangian radiative hydrodynamics \texttt{SNEC} code simulations to validate our results and obtain the luminosity in different bands, and 2D axisymmetric Eulerian hydrodynamics code \texttt{HORMONE} simulations to study the morphology of the ejecta and its deviation from spherical symmetry. The observed signature is expected to be a bright blue, UV or X-ray flare on top of the AGN luminosity from the initial shock breakout, while the subsequent red part of the lightcurve will largely be unobservable. 
We estimate the upper limit for the total event rate to be $\mathcal{R}\lesssim 100\ \rm yr^{-1}\ Gpc^{-3}$ for optimal conditions and discuss the large uncertainties in this estimate. Future high-cadence transient searches may reveal these events.  Some existing tidal disruption event candidates may originate from AGN supernovae. 
 
\end{abstract}




\section{Introduction} \label{sec:intro}


The accretion discs of active galactic nuclei (AGN) are responsible for feeding supermassive black holes (SMBHs) in the centres of galaxies \citep{lynden-bell1969}. These AGN discs are often described as geometrically thin and optically thick accretion discs, driven by the $\alpha$-viscosity \citep{ss73}. The rapidly growing observational data on AGN discs has led to more detailed models of their structure and evolution, either based on the mass flow inferred from star formation \citep{tqm05}, or on the optical/UV spectral energy distribution (SED) \citep{sg03}. 

Galactic nuclei also possess a large number of stars and stellar remnants \citep{miranda2000, bartko2010}. Some regions of the discs might become unstable and give rise to vigorous star-formation \citep{paczynski1987, dittmann2020}, and the interactions of stars and compact objects with the discs could lead to their capture and embedding into the discs \citep{artymovicz1993}. 
Stars and compact objects in a gaseous environment could accrete gas and grow over time \citep{davies2020,cantiello1}, as well as change their orbits due to the interaction with the disc \citep{Ost+83,Sye+91,artymovicz1993}. The higher stellar density in the disc, the presence of gas, and the low relative velocities may enhance the formation of binaries through tidal effects \citep{Che+99} and/or through gas-mediated capture \citep{Gol+02,Tag+20}. The interactions of binary stars with the gaseous environment could catalyze stellar mergers \citep{Bar+11}, and the formation and gravitational-wave driven mergers of compact-object binaries \citep{Mck+12}. 

Taken together, AGN disc environments could potentially increase the rate of SNe, such as core-collapse SNe or type Ia SNe. The rate of core-collapse SNe may be increased through potentially increased formation rate of massive stars \citep{artymovicz1993}. Similarly, the rate of type Ia SNe may be increased through an increased number of white-dwarf (WD) mergers \citep{mckernan2020}, or through accretion onto WDs, which can make them reach close-to-Chandrasekhar masses and explode \citep{Ost+83}. Other explosive mergers of compact objects with stars or planets could give rise to other transients such as micro - tidal disruption events (when the stars are disrupted by such compact objects, \citealp{Per+16}).  

SNe in AGNs were studied in the context of the feedback on the AGN structure and evolution \citep{Rozyczka1995, ferrara2000, morenchel2021}. Here we focus on the observational signatures of SN explosions in AGN discs. The highly relativistic electromagnetic signatures of $\gamma$-ray bursts (GRBs) and the accretion induced collapse of neutron stars (NS) into BHs have been recently explored \citep{GRB_in_AGN,  zhu2, perna20, perna2021}. Recently, \citet{zhu3} suggested that type Ia supernovae may potentially be observable in AGN discs. However, a detailed model for the interaction of a non-relativistic SN explosion with the AGN disc and its observational signatures is still lacking. In this paper, we take the first steps in the detailed modelling of general SN explosions in AGN environments.

Electromagnetic signatures from the interaction between SN ejecta and the dense circumstellar medium (CSM) have been studied intensively in the context of type IIn SNe both analytically and numerically \citep[e.g.][]{woosley2007, Chevalier2011,Ginzburg2012,Moriya2013,Dessart2015,Tsuna2019,Takei2020,Suzuki2021}. Most studies assume a wind-like dense CSM, which have density profiles following $\rho\propto1/r^2$, where $r$ is the distance from the progenitor. Some studies have also dealt with asymmetrical CSM structures, such as disc-like morphologies \citep[]{Suzuki2019}. In all cases, the main feature is that the presence of the CSM allows the kinetic energy of the ejecta to be efficiently dissipated and converted into radiation, increasing the luminosity of these events. CSM interaction is considered to be the dominant channel for producing SNe with narrow-line features (type IIn) and the extremely bright SNe called superluminous SNe \citep[SLSNe; see][for a review on SLSNe]{GalYam2019}.

\subsection{SNe in AGN discs}

Motivated by the recent success of CSM interaction models in explaining SLSNe, we develop a detailed model for the shock propagation and breakout of an SN explosion inside an AGN disc. However, there are two main differences between the wind-driven CSM SNe and our setting. While the origin of the CSM in CSM SN is the wind itself, which is already expanding  prior to the explosion \citep{nakar_piro2014, morozova_piro2017, piro2020}, in our case, the local AGN gas may be assumed to be static at the time of the explosion. The second difference is that the spatial extent of AGN discs overwhelmingly exceeds the distance scales of wind-driven explosions. For this reason, as we show further, explosions in an AGN disc can be  analogous to explosions close to the surface of a much more massive ball of gas \citep{yalinewich_matzner2019}. 




The purpose of this paper is to describe the details of SN explosions in dense environments, such as AGN discs. Here we provide an intuitive picture of the underlying physical mechanisms and key takeaway messages.

After the explosion, a shock wave is generated and propagates outward. Initially, the shock propagation velocity is much faster than the photon diffusion speed in an optically thick environment. As the shock propagates outward, the medium outside it becomes increasingly optically thin, until the photons are able to overcome the shock velocity, break out before the shock and diffuse to the photosphere. This is the breakout shell $z_{\rm bo}$, formally defined as a solution to the equation $\tau(z_{\rm bo}) = c / v(z_{\rm bo})$, where  $v(z_{\rm bo})$ is the shock velocity and $\tau(z_{\rm bo})$ is the optical depth, all evaluated at $z_{\rm bo}$. At this point, there is a surge of photons that escape out of the material, causing a sudden rise in the observable luminosity. After the initial breakout, the photosphere starts to expand and cool down, which will decrease the luminosity over time.

\subsection{Overview and structure of the paper} \label{structure}

The details of the shock propagation velocity and the breakout depend on the structure of the medium. In sec. \ref{sec:AGN}, we review the structure and properties of AGN discs and derive the vertical density profile $\rho(z)$ and the optical depth $\tau(z)$. In sec. \ref{sec:analytic}, we discuss our assumptions in \ref{sub:assumptions}, derive and compare the velocity profile $v(z)$ with numerical simulations in sec. \ref{sub:shock v}. Once  $v(z)$ is established, we find the breakout shell $z_{\rm bo}$  and the breakout time in sec. \ref{sub:bo time}. The shock velocity depends on $\rho(z)$ and $\tau(z)$, and also on the spatial location, and internal properties of the explosion, namely the explosion energy $E_0$ and the ejecta mass $M_{\rm ej}$.     

Once the breakout shell $z_{\rm bo}$ and the photosphere $z_{\rm ph}$ (defined by $\tau(z_{\rm ph})=1$) are established, the photon diffusion length scale is $d=z_{\rm ph} - z_{\rm bo}$. Since the velocity $v(z_{\rm bo})$ is also known, we can estimate the peak luminosity $L_{\rm peak}$ as the energy deposited onto the breakout shell $e_{\rm bo} \sim \rho(z_{\rm bo}) z_{\rm bo}^2 d v_{\rm bo}^2$ divided by the diffusion photon time $\sim d/v_{\rm bo}$. The luminosity is then
\begin{equation}
L_{\rm peak} = C_L \rho(z_{\rm bo}) z_{\rm bo}^2 v_{\rm bo}^3. \label{l_direct}
\end{equation} 
The dimensionless prefactor $C_L$ depends on the geometry of the explosion.

The energy deposited in layers beneath the breakout slab cannot efficiently diffuse out, and the material, therefore, cools mostly through adiabatic expansion. Adiabatic cooling leads to a rapid decay of the lightcurve: Initially, the volume is only linearly proportional with time $V\propto t$, and for a pressure $P$, the change in the energy will be $\int dE=\int P dV \propto \int V^{-\gamma} dV$. For a radiation dominated medium (which is the case after it is shocked), we have $\gamma =4/3$, and  $\int dE \propto V^{-1/3}\propto t^{-1/3}$, where we used the connection $PV^\gamma = \rm const$. The luminosity falls off as $L\propto E/t \propto t^{-4/3}$. The peak luminosity is hence dominated by the initial breakout, which is likely to be the detectable portion of the lightcurve in the luminous environment of an AGN. In dense regions where the CSM mass until the breakout is much larger than the ejecta mass $M_{\rm CSM} \gg M_{\rm ej}$, adiabatic cooling is the main source of energy loss, while for models with lighter masses $M_{\rm CSM}\lesssim M_{\rm ej}$ the optical depth is reduced, and radiative losses will also contribute, and the decay will be steeper.

After some time, the photosphere will expand further out, the volume will expand spherically, the optical depth will decrease, and deeper layers will be exposed, which makes the lightcurve decay with a much shallower slope. \cite{NakarSari2010} found a slope of $\approx -0.35$ for a radiative atmosphere profile.

The uncertainty of the peak luminosity in terms of the coefficient $C_L$ and the complex structure make it difficult to predict the peak luminosity and subsequent lightcurve. Recent detailed modelling of explosions close to a stellar surface has provided analytical expressions which were corroborated numerically \citepalias{yalinewich_matzner2019}. We extend the modelling of \citetalias{yalinewich_matzner2019} to apply it to our disc geometry under certain approximations in sec. \ref{sub:surf} and derive an approximate analytic expression for the lightcurves in sec. \ref{sub:lc}. 

In sec. \ref{sec:numerical}, we present the different initial conditions for our models, which include different internal properties and spatial locations for the explosions, different vertical profiles and SMBH masses, as well as discs with reduced densities. We also describe the numerical techniques used in this paper. We devote sec. \ref{sec:results} to describing the obtained lightcurves for the aforementioned models, both analytically and numerically. We discuss which models are in good agreement between the analytic and numerical models, and which differ and why.

Finally, we discuss the observational aspects of our results in sec. \ref{sec:discussion}. We discuss which of the events may be observable and could be distinguished from the underlying AGN luminosity. We then estimate the total rates of these events, their observed fraction and the origins of their large uncertainty. We also briefly discuss the response of the disc and comment on AGN variability and potential past and future observations of SNe explosions in AGNs. 

We make a summary of our key points in sec. \ref{sec:summary}. Readers interested in the key takeaway messages without the technical details can directly skip to this section.

\section{AGN disc structure and properties} \label{sec:AGN}
AGN discs are powered by accretion and are expected to be geometrically thin but optically thick (although there may be optically thin windows, see sec. \ref{sub:opacity}). In order to describe AGN disc structures, we use several parameters. The radial coordinate $r$ is expressed in units of the Schwarzschild radius $r_s\equiv 2GM_{\bullet} /c^2\approx 1.974M_8\ \rm AU$, where $M_8\equiv M_\bullet / 10^8\msun$ is the SMBH mass normalized to $10^8\,\msun$. The vertical structure is governed by the scale height $H$ which is generally small compared to the size of the disc ($H\ll r$). 

\subsection{Radial structure models} \label{rad}
AGN discs are generally modelled as viscous accretion discs. On the largest, galactic scales, discs are expected to be cold, gravitationally unstable and fragment into stars, with the effective Toomre $Q\equiv c_s \Omega / \pi G \Sigma \ll 1$, where $c_s$ is the sound speed, $\Omega$ is the orbital frequency and $\Sigma$ is the surface density. However, only a fraction of the gas is fragmented and the remaining gaseous component is still retained. The transition from the unstable galactic discs to accretion discs around a central object is still unclear. Angular momentum transport via global torques, rather than local viscosity, may keep the disc marginally stable and avoid fragmentation, which is the mechanism that is assumed in the  \cite{tqm05} model throughout the disc, even at smaller radii.

The structure of AGN discs can be inferred from the optical/UV spectral energy distribution (SED) in the inner regions, and from mass inflow and  star formation in the outer regions.  \citet[hereafter \citetalias{sg03}]{sg03} obtained an accretion disc model that fits the SED observations, while \cite{tqm05} constructed a model regulated by mass inflow and star formation. The SG03 model best describes the disc structure in the inner regions, up to $\lesssim 10^5 r_s$, which is the radial region of interest in our study

The radial gas density structure of the \citetalias{sg03} model may be inferred from the marginal gravitational instability condition (Eq. 15 of \citetalias{sg03}),
\begin{equation}
    \rho(r)=\frac{\Omega^2}{2\pi G Q_{\rm min}} \approx 1.22 \cdot 10^{-9} M_8^{-2} \left(\frac{r}{10^3 r_s} \right)^{-3} {\rm g\ cm^{-3}}. \label{eq:q}
\end{equation}
The latter equation is valid for $r\ge 10^3 r_s$, where $Q_{\rm min}\approx 1$ in the \citetalias{sg03} model. The disc is expected to be dominated by radiation pressure $p_{\rm rad} = aT^4/3$, where $a=4\sigma_{\rm SB}/c$ is the radiation constant and $\sigma_{\rm SB}$ is the Stefan-Boltzmann constant. In the \citetalias{sg03} model, radiation pressure indeed dominates in most parts of the disc. However, the gas pressure may have a significant contribution to the total pressure near $r\approx 10^3r_s$. Nevertheless, we use the radiation pressure throughout the paper, and the adjustments needed to include other sources of pressure are straightforward. 

The aspect ratio of a thin disc should be small, i.e. $H/r\ll1$. In the \citetalias{sg03} model, the aspect ratio is approximately given by 
\begin{equation}
    \frac{H}{r} = 8 \cdot 10^{-3} \left( \frac{r}{10^3r_s} \right)^{1/2},\label{aspect_ratio}
\end{equation}
in our range of interest.

\subsection{Vertical structure models} \label{ver}

The vertical structure of an AGN disc and its stability are also under debate (see the review of \citealp{davis_review2020} for discussion and further references). While gas pressure dominated discs are expected to be vertically stable, the stability of radiation dominated discs is undecided. Theory predicts that radiation dominated discs will be unstable \citep{ss76}. Contrary to this prediction, early radiative magneto-hydrodynamical (MHD) shearing box simulations, such as those with the ZEUS code, found that the disc is stable over many thermal timescales \citep{hirose2009}. However, these results  were later challenged by simulations with the ATHENA code. Although small box sizes reproduced the stability of \cite{hirose2009}, larger box size simulations resulted in an eventual runaway (where the disc either expands or collapses), although on much longer timescales than the thermal timescale \citep{jiang2013}. Iron bound-bound opacity may revert the situation again and make the vertical structure stable around temperatures of $1.8\cdot 10^5 {\rm K}$ \citep{jiang2016}. 

Given the uncertainties involved in the stability and the detailed vertical structure, we will remain agnostic about the true structure. Nevertheless, a different vertical structure can lead to different observational signatures for the same initial conditions of the explosion. Here we consider several possible models for the vertical structure as described below, where their main common feature is that they all describe a thin disc. Thus the vertical density profile should fall off sharply where the vertical height $h$ is larger than the scale-height, $h\gtrsim H$. Note that it is rather different from a $\rho\propto r^{-2}$ distribution that is often assumed in studies of interaction-powered SNe. For our canonical model, we assume a Gaussian density profile 
\begin{equation}
    \rho_{\rm gas}(h) = \rho_0 \exp \left(-\frac{h^2}{2H^2}\right);\quad \frac{z}{h}\in \mathbb{R}, \label{eq:rho_gas}
\end{equation}
which is applicable for gas-dominated discs and is stable. We use hydrostatic equilibrium to derive the density profile for radiation dominated discs (if they are stable) in Appendix \ref{vertical profile}, 
\begin{equation}
    \rho_{\rm rad}(h) = \rho_0 \left( 1 - \frac{h^2}{6H^2}\right)^3;\quad \left| \frac{h}{H} \right| < \sqrt{6}, \label{eq:rho_rad}
\end{equation} 
and present some of the results with this structure, as well as a simplified step profile of constant $\rho_{\rm step}=\rho_0$ up to a scale height $H$ away from the midplane. 
We assume that the disc is stable and static with the respective vertical structure prior to the explosion.  We compare the results for different vertical structures in sec. \ref{sec:disc_properties}. 

\subsection{Opacity} \label{sub:opacity}
At high temperatures, $T>10^4 \rm K$, the gas is mostly ionized and the opacity is dominated by electron scattering and bremsstrahlung (free-free). A typical value of the opacity in this range can be the electron scattering opacity $\kappa_{\rm es} = 0.2 (1+X)\ \rm cm^2\ g^{-1}$ where $X$ is the Hydrogen fraction. The opacities may be enhanced by $\approx 2$ orders of magnitude when detailed opacity tables are considered (\citealp{iglesias1996}, see also Fig. 3 of \citealp{davis_review2020}). At temperatures below $\sim 100\ \rm K$, dust absorption is the dominant effect that increases opacity, which scales as $\kappa_{\rm dust} \approx 2.4 (T/100 {\rm K})^2 \ \rm cm^2\ g^{-1}$, while the opacity is roughly constant for $T\approx 10^2-10^3\ \rm K$ in the range of $\kappa \approx 1-10\ \rm cm^2\ g^{-1}$ (see Fig 1. of \citealp{tqm05}, which is based on more recent opacity models of \citealp{semenov2003}). For intermediate temperatures around $10^3-10^4\rm K$, the gas is neutral, and most of the dust is already sublimated, thus the opacity is extremely low. Depending on the radial temperature profile, an optically thin window is expected to occur in AGN discs.  The discs are optically thin at $r\gtrsim 10^4 r_s$ in the models we adopted. Explosions in the optically thin region will look like standard SNe as the emitted radiation will be unaffected by the surrounding disc, while the associated kinetic outflows will be choked by the enormous mass. Explosions in the optically thick and massive regions ($\approx 10^4 r_s$ for $M_\bullet = 10^7 \msun$) will be completely choked (see sections \ref{sec:numerical} and \ref{sec:discussion}). For our analytical modeling, we use a constant  electron scattering opacity of solar composition ($X=0.7$), $\kappa_{\rm es}=0.34\ \rm cm^2\ g^{-1}$. We use realistic OPAL tables \citep{iglesias1996} for our radiative transfer simulations (see sec. \ref{sec:numerical} for details).

\subsection{Optical depth} \label{sub:optical depth}
For a constant opacity $\kappa$, the optical depth at height $h$, in the direction normal to the disc, is given by

\begin{equation}
\tau(h)=\kappa\intop_{h}^{\infty}\rho(h')dh'  \equiv \tau_0 \mathcal{F}(\zeta), \label{eq:tau_def}
\end{equation}
where we defined for convenience the dimensionless quantities $\tau_0 \equiv \kappa \rho_0 H$, and $\zeta \equiv h/H$. The functional form of $\mathcal{F}(\zeta) $ depends on the vertical structure.

For a uniform step density, the integral is straightforward, and the solution is $\tau(h)=\kappa\rho_{0}(H-h)=\tau_{0}(1-\zeta)$ for $0\le h\le H$ and $\tau=0$ for $h>H$ since there is no material there. This leads to $\mathcal{F}_{\rm step}(\zeta)=1-\zeta$.

For the Gaussian profile, applicable for the gas dominated disc, we have 
\begin{equation}
\mathcal{F}_{\rm gas}(\zeta) = \frac{1}{H} \intop_{h}^{\infty}\exp \left(\frac{-h'^2}{2H^{2}} \right)dh' =  \sqrt{\frac{\pi}{2}} {\rm erfc}\left(\frac{\zeta}{\sqrt{2}}\right)    
\end{equation}
where ${\rm erfc}(x)$ is the complimentary error function.

For the radiation-dominated profile, we have
\begin{equation}
\mathcal{F}_{\rm rad}(\zeta) = \frac{1}{H} \intop_{h}^{\sqrt{6}H}\left( 1-\frac{h'^2}{6H^2}\right)^3 dh' = \intop_{\zeta}^{\sqrt{6}} \left(1-\frac{\zeta'^2}{6} \right)^3d\zeta'  =  p(\zeta), \label{eq:frad} 
\end{equation}
where $p(\zeta)=16\sqrt{6}/35-\zeta+\zeta^{3}/6-\zeta^{5}/60+\zeta^{7}/1512$.

\section{Analytic lightcurve modeling} \label{sec:analytic}
Here we model the lightcurve of typical explosions. We first discuss the assumptions of the model.  We then proceed to discuss the velocity of the shock front, and finally calculate the the lightcurve by drawing analogies between our disc explosions and the recently developed theory of surface explosions. Although our analysis is generic, when applicable, we refer to our canonical model (1) of an explosion of energy $E_0=10^{51}\ \rm erg$ and ejecta mass $M_{\rm ej}=1.3 \msun$, which serves as a proxy for a standard type Ia SN explosion. The radial location of the explosion is at $r=10^3 r_s$ around an SMBH of mass $10^7 \msun$, in the disc midplane. The disc has density $\rho_0=1.22\cdot 10^{-7}\ \rm g\ cm^{-3}$ and scale height $H=8\cdot 10^{-3} r = 2.36\cdot 10^{13}\ \rm cm$ with a gas-dominated Gaussian vertical profile. For the opacity, we make use of a constant opacity $\kappa=0.34\ \rm cm^2\ g^{-1}$, which corresponds to electron scattering at Solar compositions. The optical depth is then $\tau=\sqrt{\pi/2} \kappa \rho_0 H = 1.2\cdot 10^6$. Other models are described in sec. \ref{sec:numerical} and their initial conditions are summarized in table \ref{tab:1}.

\subsection{Assumptions} \label{sub:assumptions}
Here we describe the assumptions made in constructing our solution for shock propagation, breakout and the resulting lightcurve. Consider an explosion going off in dense material. If the material is sufficiently optically thick and the prompt energy deposition is sufficiently energetic, the primary mode of energy transport, at least initially, will be kinetic outflows rather than radiation transport (photon diffusion). The initial disc temperature is $T_0$, and the explosion is spherical. For collimated outflows, not considered here, one would need to introduce an additional solid angle $\Omega_s$ spanned by the outflow. 

\textit{i) Is the outflow relativistic?}  The characteristic velocity of a non-relativistic outflow is $v_0=\sqrt{E_0/M_{\rm ej}}$, so we require that $(v_0/c)^2 = E_0/(M_{\rm ej}c^2)\ll 1$ for an outflow to be non-relativistic.  And indeed, for the canonical model, representative of type Ia SNe, $v_0/c=0.02$. Similarly, core-collapse SNe also produce non-relativistic outflows.

\textit{ii) Is the kinetic outflow the dominant form of energy transport?} For this, the velocity must be faster than the photon diffusion speed, at least initially, or, equivalently, the depth must be sufficiently large.  The optical depth at height $z$ is formally given by Eq. \ref{eq:tau_def}. If at the explosion site the local density is $\rho_0$ and the shock propagates over a typical distance,  $R_0$, the optical depth is approximately estimated as  $\tau \sim \kappa \rho_0 R_0$, where $\kappa$ is the opacity, and the diffusion speed is $c/\tau$. For kinetic outflow to dominate, we require $v_0\gg c/\tau$, or $\tau^2 E_0/M_{\rm ej}c^2\gg 1$. Combining conditions \textit{i)} and \textit{ii)}, the optical depth must also be large, $\tau \gg 1$.  Note that this defines a hierarchy of velocities $c/\tau \ll v_0 \ll c$. For our canonical model, the explosion is at the midplane, and the midplane optical depth ($z=0$) is estimated as $\tau \gtrsim 10^6$, so these conditions are initially satisfied. 

\textit{iii) Does the shock leave radiation-dominated material behind?} To answer this question, we require that the radiation energy density $a T^4$, which may be approximated by $\sim \rho v_0^2$ near shock, will be much larger than the gas energy density $\rho k_{\rm B} T/m_{\rm p}$, where $k_{\rm B}$ is Boltzmann's constant and $m_{\rm p}$ is the proton mass. Eliminating the temperature, this condition becomes equivalent to $(\rho v_0^2/a) \gg (k_{\rm B} \rho/m_{\rm p}a)^{4/3}$, or 
\begin{equation}  \label{eq:rad}
    v_0^2= \frac{E_0}{M_{\rm ej}} \gg \left( \frac{k_{\rm B}^4 \rho}{m_{\rm p}^4 a}\right)^{1/3} .
\end{equation}
 The latter in only an estimate within an order of magnitude. 

Note that the latter three conditions are equivalent to the assumptions made by  \citet[hereafter: YM19]{yalinewich_matzner2019} in the context of surface explosions. We return to the detailed modeling of \citetalias{yalinewich_matzner2019} in our lightcurve modelling is sections \ref{sub:surf}-\ref{sub:lc}. 

\subsection{Shock velocity profile} \label{sub:shock v}
Here we discuss several models for the shock velocity profile. The description is mostly generic, although sometimes, for a concrete example, we use the canonical model (model (1) in table \ref{tab:1}). We later investigate how the variation of some of the disc and explosion properties affect our results in sec. \ref{sec:results}.

For an explosion with energy $E_0$ and ejecta mass $M_{\rm ej}$, the typical velocity is $v_0 \equiv \sqrt{E_0/M_{\rm ej}}$. 
The outflow in the plane of the disc will be quenched by the disc material, and the shock will break out in the vertical direction. Therefore, we focus on the vertical height $z$ above the explosion. Generally speaking, a shock wave is always accompanied by three waves: forward shock, reverse shock (or rarefaction), and the contact discontinuity.  In our context, the forward shock compresses the CSM, while the reverse shock compresses the ejecta. The contact discontinuity separates the ejecta and the CSM matter. For most of our models, the CSM matter is more massive than the ejecta, so the reverse shock quickly traverses the ejecta while the forward shock is still propagating outwards.  We therefore focus only on the forward shock, unless stated otherwise explicitly.

The shock velocity will depend on the shock front radius $z(t)$ and on the density at this radius. Generally, three qualitatively different regimes will be evident:

\textit{i) Free expansion regime:} In this case, the swept mass $M(z)$ is much smaller than the ejected mass, i.e., $M(z) \sim \rho(z_{\rm min}) (z(t)-z_{\rm min})^3 \ll M_{\rm ej}$,  where $z_{\rm min}$ is the vertical location of the explosion site, $z(t)$ is the height reached by the shock at time $t$, and the shock velocity is close to $v_0$.
\textit{ ii) Sedov-Taylor deceleration regime:} Once the swept mass becomes comparable to the ejecta mass, $M(z)\gtrsim M_{\rm ej}$ the shock front will decelerate. Dimensional analysis leads us to a length scale $z(t)=\beta(E_{0}/\rho(z_{\rm min}))^{1/5}t^{2/5}$ and a velocity scale 
\begin{equation}
v_{\rm ST}(t)=\frac{dz}{dt}=\frac{2\beta}{5}\left( \frac{E_{0}}{\rho(z_{\rm min})}\right)^{1/5}t^{-3/5}. \label{eq:v_st}
\end{equation}

Here, $\beta$ is an order unity parameter which can in principle be estimated from the energy equation. We take for simplicity $\beta=1$. The latter scaling is the  self-similar Sedov-Taylor (ST) solution \citep{Sedov1946, Taylor1950}, where the shock front expands as $R\propto t^{2/5}$ and the shock velocity decelerates $v_{\rm ST} \propto t^{-3/5}$.

\textit{iii) Sakurai accelerating regime:} The ST picture is correct if the density is uniform.  The vertical structure of the disc is generally far from uniform. Close to the disc edge (either the physical edge or the photosphere, if the disc is formally infinite), the density gradient is quite steep, and the shock propagates more easily, which causes it to accelerate with decreasing density as $v\propto\rho^{-\mu}$, which is known as the Sakurai law \citep{Sakurai1960}. For radiation dominated material, $\mu \approx 0.19$ fits well with previous estimates (see also \citealp{MM99}, hereafter \citetalias{MM99}, their sec. 4.2 and references therein for more details and discussion). Note that the latter picture is valid if the swept mass $M(z)\gtrsim M_{\rm ej}$. Some of our models have the swept mass comparable to or smaller than $M_{\rm ej}$, and in these cases the ST phase is skipped and the shock may only accelerate or directly break out.

Close to the centre, the density is almost constant, and the evolution is as in the ST regime. Close to the edge, a small change in the radial scale leads to a large change in the density, and therefore the shock will follow the Sakurai law. It is possible to write down a general formula that includes both cases, as done in \citetalias{MM99}:

\begin{equation}
v(t)=v_{0}\left(\frac{t_{1}}{t}\right)^{3/5}\left(\frac{\rho(z(t))}{\rho(z_{\rm min})}\right)^{-\mu},\label{v_extra}
\end{equation}
where $t_1$ is the time when a transition from the free expansion to the ST deceleration occurs and is  given by $v_{\rm ST}(t_1)=v_0$ from Eq. \ref{eq:v_st}. Plugging $t_1$ into the ST length scale leads to $z(t_1)=z_1=(4M_{\rm ej}/25\rho(z_{\rm min}))^{1/3}$. The transition occurs when the swept mass is $4M_{\rm ej}/25=0.16M_{\rm ej}$. For the ejecta mass of $1.3 \msun$, as in our model 1, the transition occurs at $0.21 \msun$, or at radius $z\approx 1.5 \cdot 10^{13}\ \mathrm{cm} =0.63 H$.

The transition from the ST decelerating velocity to Sakurai accelerating velocity occurs when $v(t)$ is at a minimum. From the ST solution, we can invert $z(t)$ to $t(z)\propto z^{5/2}$ and find the location of the minimum $z_2$ given by $d\ln\rho/d\ln z|_{z_2}=-3/(2 \mu)$. For a Gaussian profile, the velocity behaves as $v\propto z^{-3/2}\exp(\mu z^2/(2H^2))$, the logarithmic derivative of the density is  $d\ln\rho/d\ln z=-(z/H)^2$, and  the  minimum is achieved at $z_2/H=(3/(2\mu))^{1/2}\approx 2.8$. 

The three regimes and their transitions described above allow us to construct two possible solutions.  The first is the piecewise velocity solution.  Here we divide the behaviour into three regimes, where we first begin with free expansion and transform to ST deceleration. The ST approach is not accurate since it was developed for a uniform density. The density change is negligible for $z\ll H$, but it is already significant at $z_1$. We use an effective scaling where we replace $\rho(z_{\rm min})$ in the ST solution by $\rho(z)$. Although not entirely self-consistent, we can write $v$ as in Eq. (\ref{v_extra}) but with an effective density power law $\mu'=\mu+1/5$, where the $1/5$ is coming from the ST scaling of $z\propto \rho^{-1/5}$. The piecewise velocity is 
\begin{align} \label{vpi}
    {v_{\rm p}(z)=v_{0}\begin{cases}
1 & z<z_{1}\\
\left(\frac{z_{1}}{z}\right)^{3/2}\left[\frac{\rho(z)}{\rho_{1}}\right]^{-(\mu+1/5)} & z_{1}<z<z_{2}\\
\left(\frac{z_{1}}{z}\right)^{3/2}\left(\frac{\rho_{2}}{\rho_{1}}\right)^{-(\mu+1/5)}\left[\frac{\rho(z)}{\rho_{2}}\right]^{-\mu} & z>z_{2} 
\end{cases}} 
\end{align}
where $\rho_i=\rho(z_i)$ for $i=1,2$.

Another solution is the direct \citetalias{MM99} shock velocity 
\begin{equation}
 v_{\rm MM99}(z)=\left(\frac{E_{0}}{M_{{\rm ej}}+M(z)}\right)^{1/2}\left(\frac{\rho(z)}{\rho(z_{\rm min})}\right)^{-\mu}, \label{vMM99}
\end{equation}
 Note that Eq. (\ref{vMM99}) captures well all of the regimes. For $M_{\rm ej}\gg M(z)$ the density and hence the velocity are essentially constant. When $M(z)\gg M_{\rm ej}$ but $\rho(z) \sim \rho(z_{\rm min})$, the velocity is $\propto z^{-3/2}\propto t^{-3/5}$ as expected from a ST solution.  Finally, once $\rho(z)\ll \rho(z_{\rm min})$, the Sakurai term dominates the acceleration.
 
 \begin{figure}
     \centering
     \includegraphics[width=8cm]{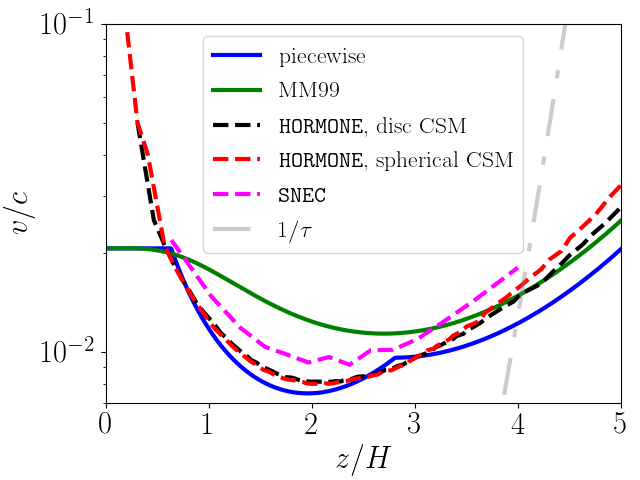}
     \caption{Shock velocity prescriptions for model 1 in the AGN disc, in units of the speed of light $c$. The solid lines are the theoretical shock velocities; the piecewise model (blue), Eq. (\ref{vpi}), and the \citetalias{MM99} model (green), Eq. (\ref{vMM99}). The dashed lines are the numerical shock velocities. The hydrodynamical code \texttt{HORMONE} is used to obtain the matter velocity $v_m$ just behind the shocked CSM, and the shock velocity is $v_m (\gamma+1)/2$ in the disc geometry (dashed black) and spherical geometry (dashed red). The result from the spherically-symmetric radiative transfer code \texttt{SNEC} is also presented in magenta. For completeness, the photon diffusion velocity, which is also the inverse of the optical depth, is shown in dot-dashed grey lines. The shock breakout occurs when the gray line crosses the shock velocity, which occurs close to $z\sim 4 H$ for all shock velocity estimates for model 1. }
     
     \label{fig:velocity_models}
 \end{figure}
 
Figure \ref{fig:velocity_models} shows the different velocity models for the canonical (model 1) explosion together with the shock velocity obtained numerically. The detailed description of the hydrodynamical code \texttt{HORMONE} and the radiative transfer code \texttt{SNEC} appears in sec. \ref{sec:numerical}. The three velocity regimes are clearly evident. The flat free expansion phase lasts until $z_1/H\approx 0.63$ for model 1, where the velocity starts to decelerate. The transition to Sakurai acceleration occurs somewhere around $z/H\sim 2-2.8$. We see that the matter-derived shock velocity $v_m(\gamma+1)/2$ better agrees with the piecewise model, while the \texttt{SNEC} shock velocity is somewhere in between the piecewise and \citetalias{MM99} model. In either case, the overall acceleration in the low-density outskirts of the disc is underestimated.

In summary, there are two main theoretical shock velocity estimates for an explosion inside the AGN disc: a piecewise fit explicitly taking account the three distinct propagation regimes (Eq. \ref{vpi}), and a direct continuous measure of the swept mass (\citetalias{MM99} - Eq. \ref{vMM99}). The numerical shock velocity falls between the two theoretical shock velocity models.

\subsection{Breakout time}\label{sub:bo time}
Even though the shock is accelerating in the Sakurai regime, the photon diffusion velocity rises more steeply. Eventually, the photon diffusion speed $c / \tau(z)$ will intersect the shock velocity $v(z)$. The two velocities are equal when the grey dot-dashed line intersects the velocity model of choice in Fig. \ref{fig:velocity_models}. The location of this intersection defines the breakout shell $z_{\rm bo}$, which is implicitly given by $\tau(z_{\rm bo}) = c /v(z_{\rm bo})$ and is around $\sim 4 H$ for model 1, regardless of the velocity model.

The breakout time $t_{\rm bo}$ is the time elapsed for the first light to be beyond the breakout shell from the time of the explosion. In order to finally escape, the photons need to diffuse up to the photosphere. This diffusion time is much shorter than the breakout time.

The breakout time can be calculated by 
\begin{equation}
    t_{\rm bo} = \intop_{z_{\rm min}}^{z_{\rm bo}} \frac{dz}{v(z)} \label{eq:t_bo}
\end{equation}

The exact expressions are given either by special functions or integrated numerically. We provide them in Appendix \ref{appendix-breakout}, together with the expressions for off-plane explosions. 

\subsection{Modelling AGN explosions as surface explosions} \label{sub:surf}

Surface explosions are explosions that occur in a relatively sparse material, close to a surface of a medium and far away from the dense core of the medium (e.g. a massive star). \citetalias{yalinewich_matzner2019} studied in detail such surface explosion in massive stellar envelopes. Our explosion is analogous to a surface explosion in the following way. Due to the planar geometry of the disc and its large spatial extent $r \gg H$, an explosion in the midplane of the disc  cannot propagate in the in-plane direction, and the breakout occurs effectively on the vertical surface of the disc. This is analogous to an explosion at a distance $l\ll R_{\star}$ below the surface of a star, the model for which was recently put forward by \citetalias{yalinewich_matzner2019}. The shock propagates only upwards. Here we parametrise our explosion as a surface explosion, with some limitations.

\subsubsection{Choosing length scales} \label{choose_leff}
The \citetalias{yalinewich_matzner2019} model assumes that the density vanishes at some boundary $z_b$ and that the density scales as a power law profile of the form $\rho = \rho_{0}(x/l)^{\omega}$ where $x$ is the \textit{distance from the edge} and $l$ is a typical length scale of order the distance of the explosion hotspot from the edge. When applied to our extended model, $l$ is generally on the order of H, but its exact definition has some ambiguities as we discuss below.

The surface explosion modelling in \citetalias{yalinewich_matzner2019} is encapsulated by two dimensionless parameters,  $\Gamma\equiv E_{0}/(c^{2}\rho_{0}l^{3)}$ and $\tau_{0}=\kappa\rho_{0}l$. Initially, the only length scale we have is $H$. In \citetalias{yalinewich_matzner2019}, $l$ has several different roles:

\textit{i) The distance from the explosion to the edge.}

\textit{ii) The transition between the end of the Sedov-Taylor phase and the beginning of the shock acceleration phase.}

\textit{iii) The value in the expression for the velocity at the end of the Sedov-Taylor phase, $v_{{\rm ST}}=\sqrt{E_{0}/\rho_{0}l^{3}}$.}

These three definitions appear to be contradictory. First, the free expansion phase is ignored, or a  point-source explosion is assumed. Second, after passing a length $l$ (as in i.), formally, we should be at the edge, which is definitely after the  breakout. However, according to ii. and iii., it is only the phase where the shock starts to accelerate and is assumed to be optically thick.

These differences are small in the stellar atmosphere with its steep density profile towards the edge. However, we saw that for model 1, the differences between the transition to accelerating shock at $z_{2}=2.81H$, the breakout shell at $z_{{\rm bo}}=3.99H$ for the piecewise model and $z_{{\rm bo}}=3.96H$ for the \citetalias{MM99} model, and the photosphere at $z_{{\rm ph}}=4.93H$, are around a scale height away from each other. Moreover, free expansion phase ends at $z_{1}=0.64H$, which is also comparable to the scale height. 

The velocity at the end of the ST phase is $v_{\rm ST} \sim c\Gamma^{1/2}$. Hence the effective length scale $l_{\rm eff}$ in this case should replicate the total range of the swept mass from the end of the free expansion phase until the breakout shell. We therefore set $l_{\rm eff}=z_{\rm bo} -z_1$, unless otherwise specified. 
\begin{equation}
    \Gamma = \frac{E_{0}}{c^{2}(M_{{\rm ej}}+\rho_{0}l_{{\rm eff}}^{3})}. \label{Gamma}
\end{equation}
Here, the inclusion of $M_{\rm ej}$ in the denominator extends the analysis of \citetalias{yalinewich_matzner2019} to include the free expansion phase and also avoids unphysically large values of $\Gamma$ for low densities or small $l_{\rm eff}$. 

The optical depth $\tau_0$ should be comparable with the midplane optical depth, hence we keep $\tau_0$ as $\kappa \rho_0 H$.
\subsubsection{Effective power law index $\omega_{\rm eff}$ }
Once we set $l_{\rm eff}$ and $\Gamma$ for the explosion and also have $\rho_0, H$ and $\tau_0$ from the AGN disc model, we can assign an effective power law index $\omega_{\rm eff}$ in the following way:  $\omega_{\rm eff}$ is a local value and formally varies along the vertical shock propagation. 

For a local density profile 
\begin{equation}
    \rho(x)=\rho_a (x/l_{\rm eff})^{\omega_{\rm eff}},  \label{eq:omega}
\end{equation}
where $\rho_a$ is the density at a distance $l_{\rm eff}$ from the edge, and $0\le x \le l_{\rm eff}$ is measured from the edge. 
We impose a matching condition on $d\ln\rho/dz$ between the assumed disc density profile and the power law model in order to fix $\omega$:

\begin{equation}
    \rho_{a}=\rho(z_{{\rm ph}}-l_{\rm eff});\quad\omega=-l_{\rm eff}\left.\frac{d\ln\rho}{dz}\right|_{z_{\rm ph}-l_{\rm eff}}\ . \label{omega_eff_gen}
\end{equation}
For the slab profile, the derivative is zero and thus $\omega_{\rm eff}=0$ everywhere. For the Gaussian profile,
$d\ln\rho/dz=-z/H^{2}$, and 
\begin{equation}
\omega_{{\rm gas}}(l_{{\rm eff}})=\frac{l_{{\rm eff}}(z_{\rm ph}-l_{{\rm eff}})}{H^{2}}. \label{omega_eff_gas}
\end{equation}
For the radiation dominated profile the power law is
\begin{equation}
\omega_{{\rm rad}}(l_{{\rm eff}})=6\frac{\sqrt{6}H-l_{{\rm eff}}}{2\sqrt{6}H-l_{{\rm eff}}}, \label{omega_eff_rad}
\end{equation}
where we used the actual edge of the disc $z_{\rm edge}=\sqrt{6} H$ instead of the photosphere.

We note that $\rho_a$, $l_{\rm eff}$ and $\Gamma$ are determined by the extent of the mass swept by the outflow prior to the breakout. $\tau_0$ is also well defined. Although Eq. \ref{omega_eff_gen} suggests that determining $l_{\rm eff}$ uniquely determines $\omega_{\rm eff}$, this is misleading, since in the \citetalias{yalinewich_matzner2019} model of Eq.~(\ref{eq:omega}), the density profile is a global power with index $\omega_{\rm eff}$, while in our case the power law is only a local fit to the density profile. Therefore, we have the freedom to decide where to match the density profile to the power law.  We opted to do so at the depth of $l_{\rm eff}$ beneath the breakout shell in equations \ref{omega_eff_gas} and \ref{omega_eff_rad}.  We could alternatively choose to do so at the breakout radius $z_\mathrm{bo}$ in an attempt to model the immediate post-breakout behavior more accurately; we consider this alternative choice in section \ref{sec:omega}.
 Regardless of the local choice of $\omega_{\rm eff}$, the dependence of the luminosity on $\omega$ is expected to be much weaker than on the other parameters.

\subsection{Lightcurves} \label{sub:lc}

As mentioned in \ref{structure}, the slab geometry leads to adiabatic expansion in the optically thick regime and rapid decay. The setting is similar to an explosion close to a stellar surface studied by \citetalias{yalinewich_matzner2019}. They applied the latter arguments for a surface explosion using the dimensionless parameters $\Gamma$ and $\tau_0$ as discussed in sec. \ref{sub:surf}, where an additional phase, where the hotspot of the explosion becomes optically thin enough, which allows depletion of material deeper than the hotspot and forming a crater. The time of the transition where the hotspot is exposed is found to be $t_{\rm sph}=(H/c) \tau_0^{1/2} \Gamma^{-1/4}$ (Eq. 24 of \citetalias{yalinewich_matzner2019}). 

Once we find $l_{\rm eff}$, $\omega_{\rm eff}$, followed by $\Gamma$ and $\tau_0$ as describe in sec. \ref{sub:surf}, we can use \citetalias{yalinewich_matzner2019}'s modelling for the lightcurve. From now until the end of the section, we drop the 'eff' subscript and simply write $\omega$ to avoid cumbersome notation. The resulting luminosity is 

\begin{equation}
\frac{L(t)}{E_0c/H}= 
\begin{cases}
\Gamma^{(\omega\mu -2\omega/3 - 5/6)\delta_{-} } \tau_0^{(5\omega\mu/3 - \omega -4/3)\delta_{-}} \left( \frac{ct}{H} \right)^{-4/3} & \rm{pl}\\
\Gamma^{(-\omega\mu + 1/6)\delta_{+} } \tau_0^{(\omega\mu/ - \omega -4/3)\delta_{+}} \left( \frac{ct}{H} \right)^{(-4\omega\mu+2/3)\delta_{+}} & \rm{sph}
\end{cases}
\label{lightcurve}    
\end{equation}
in the planar and spherical phases, respectively, where $\delta_{\pm} \equiv (1\pm \omega\mu + \omega)^{-1}$.  For concreteness, we choose $\omega(l_{\rm eff})$ as in equations \ref{omega_eff_gas} and \ref{omega_eff_rad}. 

 %
In certain cases, we will use an expansion of $\omega_{\rm eff}$ close to the edge, where we use $\omega^{\rm edge}_{\rm eff} = \omega_{\rm eff}(l_{\rm edge})$ with $l_{\rm edge} = z_{\rm edge} - z_{\rm bo}$, where the edge is either the physical edge of the disc (for radiative and slab models), or the photosphere  (for Gaussian models). We discuss this choice and its implications in the results section.


\begin{figure}
    \centering
    \includegraphics[width=8cm]{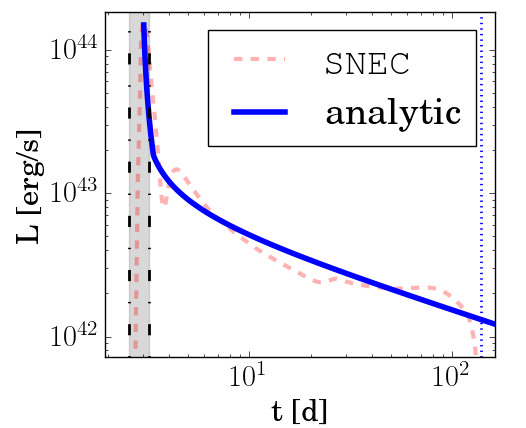}
    \caption{Analytical lightcurve for model 1. We show the analytic bolometric luminosity (i.e. the lightcurve) as a function of time with the solid blue curve. The numerical \texttt{SNEC} result is shown as a  dashed red curve. The analytic curve is the maximum of the planar and spherical phases in Eq. \ref{lightcurve}. The vertical dot-dashed lines are the breakout times for the two velocity models   given in Eq. \ref{t_pw} and \ref{t_mm99}, where the rightmost one is for the piecewise model since it has a lower velocity than the \citetalias{MM99} model. The gray area between them is the expected range of the breakout times. The dotted blue vertical line is the end of the spherical expansion phase,  $t_{\rm sph}$.}
    \label{fig:analytica_model1}
\end{figure}

Fig. \ref{fig:analytica_model1} shows the analytic lightcurve for model 1 in solid blue and the numerical \texttt{SNEC} lightcurve in dashed red. We shifted the analytical curve in time so that the breakout falls on top of the maximal luminosity of the numerical curve.  After the peak, \texttt{SNEC} model  luminosity is initially more shallow than the $L \propto t^{-4/3}$ decay expected in the planar phase from the simple arguments of section \ref{structure}, but subsequently steepens before becoming more shallow again at the transition to the spherical phase. After the initial breakout with $L \sim 10^{44}\ \rm erg\ s^{-1}$, the luminosity rapidly decays by one order of magnitude over $\sim 0.4\ \rm d$ from breakout and then decays more shallowly during the spherical phase. The end of the spherical phase is indicated by the dotted vertical line around $\sim 150\ \rm d$. The gray area in the region between $t_{\rm bo}|_{\rm MM99} < t < t_{\rm bo}|_{\rm pw}$ is given by Eqs. \ref{t_mm99} and \ref{t_pw}, respectively, and indicates the range of times when the breakout occurs. In addition, the \texttt{SNEC} lightcurve is also plotted in dashed red, where the peak luminosity of the analytic model is set to match the peak luminosity of the \texttt{SNEC} lightcurve, which falls in the margins of the breakout time. Overall, there is a good correspondence, and the bumps in the \texttt{SNEC} lightcurve can be attributed to the reverse shock initially and varying opacity effects at later times, as discussed in detail in section \ref{sec:results}.

We note that in order to set $z_{\rm bo}$ (and then $l_{\rm eff}$, $\omega$, $\Gamma$ and the lightcurve), we need to solve for the breakout shell $\tau(z_{\rm bo}) = c/v_{\rm bo}$. This is done numerically using the standard \texttt{fsolve} function of \texttt{Python}'s \texttt{scipy.optimize}  package. We also need to choose a model for the velocity. We use the earlier derived piecewise model, Eq. \ref{vpi}, since the numerical breakout time is close to the breakout time evaluated using this velocity prescription. We also tried the \citetalias{MM99} prescription, and it gave almost indistinguishable results in most cases. 

\section{Numerical Simulations} \label{sec:numerical}

Here we describe the numerical approach and initial conditions for our simulations and analysis.

\subsection{Initial conditions}

We already introduced the initial conditions for our baseline model at the beginning of section \ref{sec:analytic}. In order to explore the sensitivity to other parameters of the disc and the explosion, we also studied several other models, varying one or two parameters at a time. For changes in the global disc structure with the SMBH mass, we adopt the following scalings from the \citetalias{sg03} model. At radial locations scaled by the gravitational radius of the SMBH, i.e. at $r\propto r_s \propto M_\bullet$, the midplane density is $\rho \propto \Omega^2 \propto M_\bullet/r^3 \propto M_\bullet^{-2}$. The rescaled interacting mass is $\rho H^3 \propto M_\bullet$.  The optical depth is also rescaled as $\tau \propto \rho H \propto M_\bullet^{-1}$. The aspect ratio is taken from Eq. \ref{aspect_ratio}.



\begin{table}
\begin{center}
\begin{tabular}{|c|c|c|c|c|c|c|c|}
\hline 
\# & $M_\bullet$ & $r/r_{s}$ & $E_{51}$ & $\rho(z)$ & $z/H$ & $M_{\rm CSM}/\msun$ & Remarks\tabularnewline
\hline 
\hline 
1 & $10^{7}$ & $10^{3}$ & $1$ & Gas & $0$ & $12.8$ &  a\tabularnewline
\hline 
2 & $10^{7}$ & $10^{4}$ & $1$ & Gas & $0$ & $390$ & b\tabularnewline
\hline 
3 & $10^{8}$ & $10^{3}$ & $1$ & Gas & $0$ & $128$ &c\tabularnewline
\hline 
4 & $10^{6}$ & $10^{4}$ & 1 & Gas & 0 & $39$ & b,c\tabularnewline
\hline 
5 & $10^{7}$ & $10^{3}$ & $10$ & Gas & $0$ & $12.8$ & d\tabularnewline
\hline 
6 & $10^{7}$ & $10^{3}$ & $0.1$ & Gas & $0$ & $12.8$ & d\tabularnewline
\hline 
7 & $10^{7}$ & $10^{3}$ & $1$ & Rad & $0$ & $7.6$ &e\tabularnewline
\hline 
8 & $10^{7}$ & $10^{3}$ & $1$ & Step & $0$ & $3.4$ & e\tabularnewline
\hline 
9 & $10^{7}$ & $10^{3}$ & $1$ & Gas & $1$ & $1.92$ &f\tabularnewline
\hline 
10 & $10^{7}$ & $10^{3}$ & $1$ & Gas & $2$ & $0.15$ & f\tabularnewline
\hline 
11 & $10^{7}$ & $10^{3}$ & $1$ & Rad & $1$ & $0.5$ &e,f\tabularnewline
\hline 
12 & $10^{7}$ & $10^{3}$ & $1$ & Rad & $2$ & $6.4\cdot10^{-4}$ &e,f\tabularnewline
\hline 
13 & $10^{6}$ & $10^{3}$ & 1 & Gas & 0 & $1.28$ &c \tabularnewline
\hline 
14 & $10^{7}$ & $10^{3}$ & $1$ & Gas & $0$ & $1.28$ &g\tabularnewline
\hline
15 & $10^{7}$ & $10^{3}$ & $1$ & Gas & $0$ & $12.8$ &h\tabularnewline
\tabularnewline
\end{tabular}
\caption{Initial conditions for the simulations. The columns are: simulation number, central BH mass (in $M_{\odot}$), radial location, explosion energy (in $10^{51}\rm \ erg$), the vertical height above the midplane, CSM mass (Eq.~\ref{eq:m_int}), and remarks describing the parameters being varied: a) The canonical model. b) Model with varied radial distance $r$. c) Model with varied mass of the SMBH. d) Model with varied explosion energy $E_0$. e) Model with varied vertical density profile $\rho(z)$. f) Model with varied vertical distance $z$. g) Model with a reduced density $\rho_0 = 1.23\cdot 10^{-8}\ \rm g\ cm^{-3}$ to mimic a starved AGN. h) Model with a larger ejecta mass, $M_{\rm ej} = 10 \msun$, to mimic a core-collapse SN.} \label{tab:1}
\par\end{center}
\end{table}

Table \ref{tab:1} lists the grid of initial conditions for the numerical simulations. Model 1 is our baseline canonical model, representing a typical type Ia SN explosion. Models 2-4 and 13 explore the impact of the SMBH masses and the radial location. Models 5-6 explore the impact of the explosion energy. Models 7-8 and 11-12 explore the impact of the vertical profile. Models 9-12 explore the impact of the vertical location. Model 14 mimics a starved AGN disc with reduced density, while Model 15 mimics a core collapse SN by considering more massive ejecta than in other models, $M_{\rm ej}=10\ M_{\odot}$. The explosion energy of SN 1987A is estimated at $(1.5\pm0.12) \times 10^{51}$ erg \citep{1987A}. The observed explosion energies of core-collapse SNe range from a few times $10^{50}$ erg to $\sim 10^{53}$ erg for hypernovae with massive progenitors \citep{janka2012}. We therefore keep the explosion energy at $10^{51}$ erg for model 15 as most representative for core-collapse SNe.
 
The CSM mass in the penultimate column of table~\ref{tab:1} is defined as 
\begin{equation}
M_{\rm CSM} =4\pi\intop_{z_{\rm min}}^{z_{\rm max}}\rho(z)(z-z_{\rm min})^{2}dz, \label{eq:m_int}
\end{equation}
which depends on the density profile, and the maximal and minimal vertical locations. In Eq. \ref{eq:m_int}, we assumed a spherical approximation, namely that the most significant interaction of the ejecta with the disc material will be in the vertical direction, allowing us to approximate the density profile at radius $r$ as $\rho(r)=\rho(z=r)$. This is true in our geometry since the in-plane directions are choked. We perform the explicit calculation of the CSM mass for each model in Appendix \ref{app:int_mass}. 
 
\subsection{Spherically symmetric lightcurve modelling}

To model the lightcurves arising from the CSM interaction numerically and verify our analytic models, we perform detailed simulations with the supernova explosion code \texttt{SNEC} \citep{snec_code, snec_code2}. \texttt{SNEC} is a spherically-symmetric Lagrangian radiative-hydrodynamics code that accounts for shock propagation through the use of artificial viscosity. Compared to our analytic models, \texttt{SNEC} directly solves the equations of radiative hydrodynamics and, therefore, allows us to assess the validity of our approximations. Additionally, it allows us to gain some intuition into the importance of ionisation, detailed opacities and heating from $^{56}{\rm Ni}$ for forming the final lightcurves. 

\texttt{SNEC} uses OPAL opacities \citep{iglesias1996} in the high-temperature regime (at $3.75<\log_{10} T/{\rm K}<8.7$) and the solar-scaled \citet{ferguson2005} opacities, which accounts for molecular contributions, in the low-temperature regime (at $2.7<\log_{10} T/{\rm K}<4.5$). In the overlapping regions, the low-temperature opacity is preferred. 
The code accounts for ionisation states by solving the equilibrium Saha equation and models radiative transfer through flux-limited equilibrium photon diffusion. \texttt{SNEC} also models the contribution to the lightcurve from the radioactive decay of $^{56}{\rm Ni}$ and $^{56}${\rm Co}. The decay leads to the deposition of gamma-rays and positrons, which heat the ejecta, with gamma-rays dominating the heat deposition. The code solves for gamma-ray propagation and absorption through solving for radiative transfer equation with grey opacities. 


We set up explosions by initialising balls of material with an exponential density profile with a radial scale height set so that the density at the boundary of the ball is $1\,{\rm g}/{\rm cm}^3$. Such a setup allows for a mild density contrast near the CSM boundary, which improves the resolution of the shock. For models 1-14, having a type Ia-like engine, we set the ball radius to $0.1 \rsun$ and assign it $1.3 \msun$ of C-O material. For model 15, we set a $10 \msun$ ball of $2 \rsun$ radius, thus mimicking a core-collapse SN potentially stripped by a companion. The engine acquires the explosion energy split equally into kinetic energy and thermal energy. Furthermore, the kinetic energy is distributed to provide comparable escape velocities for all the velocity bins of ejecta. We have verified that details of the setup, such as the fraction of the explosion energy injected as thermal energy or the initial radius of the explosion ball, have a negligible effect on the lightcurves. We simulate all the models twice, with and without $^{56}$Ni. To compare our results to the analytic models of CSM interaction, we use the simulations without  $^{56}$Ni, and we examine the qualitative effects of $^{56}$Ni in the results section~\ref{sec:results}. For models with $^{56}$Ni, we assume $0.6\,\msun$ of the radioactive material, typical for type Ia explosions (apart from model 15, wherein we assumed $0.2\,\msun$ of $^{56}$Ni). The CSM is set up in concentric shells, following the density and composition profiles for the different AGN models, with the full models containing between $1000$ and $2000$ grid cells. We sample the lightcurves produced in \texttt{SNEC} with a step of $5$ minutes to resolve the sharp hour-long peaks in some of the models. Overall, the spherically-symmetric models are initialised similarly to how it is done in the analytic modelling.

The code allows us to trace the detailed properties of the CSM and the shock over time and calculate the lightcurves for the models.

\subsection{Effects of the geometry on hydrodynamics and lightcurves}

While the \texttt{SNEC} code can deal with most of the relevant physics such as radiative transfer, detailed opacity tables, nuclear energy deposition from $^{56}$Ni and ionisation, it cannot model non-spherical geometries. To understand how the supernova ejecta interact with the non-spherical AGN disc material, we carry out additional 2D axisymmetric hydrodynamic simulations of the interaction. For this, we use the hydrodynamic code \texttt{HORMONE}, which is a grid-based code that solves the hydrodynamic equations through a Godunov-type scheme \citep[]{hirai2016}. We employ an equation of state with contributions from ideal gas and radiation. A spherical coordinate system is used, where we assume axisymmetry with the symmetry axis taken perpendicular to the disc and the coordinate origin placed at the centre of the supernova explosion. In our axisymmetric treatment we ignore any motion of gas in the disc prior to the explosion, such as shearing motion expected in Keplerian discs, and instead treat the disc as an initially static slab of material. We also could not apply the gravity from the central SMBH, so we ignore all gravitational forces, including self-gravity. Therefore, we neglect the initial thermal energy in the disc material too, since it will lead to artificial expansion without the gravitational forces that keep it bound. These assumptions are still valid as long as the time-scale for the shock to reach the surface of the disc is much shorter than the local Keplerian period, and the shock energy is larger than the pre-explosion thermal energy of the disc.

We set up the simulation by placing a slab of material to represent the AGN disc. At the coordinate origin, we place a homologously expanding supernova ejecta model with an exponential density profile as
\begin{equation}
 \rho_\textnormal{ej}(r)=\frac{3\sqrt{6}M_\textnormal{ej}}{4\pi v_0^3t_0^3}\exp{\left(-\frac{\sqrt{6}r}{v_0t_0}\right)},
\end{equation}
where $t_0$ is the time since explosion. 
Such exponential profiles are commonly seen in type Ia SN ejecta models  \citep[e.g.][]{nomoto1984}. The velocity profile is set to be $v(r)=r/t_0$. When integrated from 0 to infinity, the total mass is $\int_0^\infty4\pi r^2\rho_\mathrm{ej}dr=M_\mathrm{ej}$ and the total kinetic energy becomes $\int_0^\infty\frac12\rho_\mathrm{ej}(r)v(r)^2\cdot4\pi r^2dr=E_0$. We cut off the explosion profile at a radius $r=R_\textnormal{exp}$ and the time since explosion is set through $t_0=R_\textnormal{exp}/(4v_0)$. The factor 4 is chosen so that the total integrated ejecta mass and energy are still very close to $M_\mathrm{ej}$ and $E_0$. We set a small enough value of $R_\textnormal{exp}$ such that the mass of the disc material within that radius is negligible compared to the ejecta mass ($\int_0^{R_\textnormal{exp}}4\pi r^2\rho_\textnormal{disk}dr\ll M_\textnormal{exp}$), where we can safely assume that the ejecta have freely expanded up to that radius. For all the models presented here, we use an ejecta mass of $M_\textnormal{ej}=1.3~\msun$, which represents the ejecta mass for typical type Ia SNe, except for model 15, which has a representative mass of $M_{\rm ej}=10 \msun$, typical for core-collapse SNe.

The ejecta are covered with at least 30 grid points, and the radial grid size is increased as a geometrical series as it goes out. The outer boundary is taken at $\sim100~H$, and we use $\sim1000$ grid points in the radial direction. We divide the polar direction into 400 cells spaced uniformly in $\cos\theta$ so that the solid angle of each cell is equal and the disc is properly resolved.

The lightcurves are estimated through a post-process ray-tracing procedure similar to that of \citet[]{suzuki2016}. See Appendix~\ref{app:lightcurve_hormone} for more details of the lightcurve calculation method.

\section{Results} \label{sec:results}

Here we present the analytic and numerical lightcurves for the different models. 

\subsection {Peak luminosities}

\begin{table}
\begin{center}
\begin{tabular}{|c|c|c|c|c|}
\hline 
 & $L_{44}^{{\rm YM19}}$ & $L_{44}^{{\rm SNEC}}$ & $C_L^{{\rm YM19}}$  & $C_L^{{\rm SNEC}}$\tabularnewline
\hline 
\hline 
1 & 1.48 & 1.7 & 7.97 & 9.12\tabularnewline
\hline 
2 & 0.12  & 0.14 & 5.54 & <6.48\tabularnewline
\hline 
3 & 0.26  & 0.28 & 4.34 & 4.69\tabularnewline
\hline 
4 & 0.62  & 0.93 & 5.8 & 8.6\tabularnewline
\hline 
5 & 29.38 &  27.68 & 10 & 9.44\tabularnewline
\hline 
6 & 0.08  & 0.13 & 6.63 & 10.87\tabularnewline
\hline 
7 & $4.24^a$ & 4.42 & 6.91 & 7.21\tabularnewline
\hline 
8 & $24.70^b$  & 33.5 & 1.15 & 1.56\tabularnewline
\hline 
9 & $3.96^a$  & 7.86 & 10.65 & 30.38\tabularnewline
\hline 
10 & $12.77^a$ &  14.90 & 40.97 & 26.84\tabularnewline
\hline 
11 & $7.82^a$ & 21.93 & 10.68 & 26.82\tabularnewline
\hline 
12 & $22.55^a$ &  5.43 & 47.2 & 11.53\tabularnewline
\hline 
13 & 7.85  & 2.65 & 21.05 & 7.1\tabularnewline
\hline 
14 & 8.89  & 14.16 & 14.69 & 23.4\tabularnewline
\hline 
15 & 1.45  & 1.51  & 13.45 & 14.04\tabularnewline
\hline\tabularnewline
\tabularnewline
\end{tabular}
\caption{Peak luminosities and their coefficients. The first column is the model number. The luminosities are normalized to $L_{44} = L/(10^{44}$ erg/s). The second column is $L^{\rm YM19}$, the peak luminosity derived from the maximal value of Eq. \ref{lightcurve}. The third column is the maximum luminosity obtained in \texttt{SNEC} simulation. The forth column is the ratio $L^{\rm YM19}/(\rho z^2 v^3)_{\rm bo}$, which is an estimate for $C_L$ from Eq. \ref{l_direct}. The denominator is evaluated at the breakout shell. The last column is the same, only for the numerical maximal \texttt{SNEC} luminosity. a) Evaluated $\omega$ at the breakout radius for the analytical fit. b) Used $\omega=1$. }  \label{tab:2}
\par\end{center}
\end{table}

In table \ref{tab:2}, we show the peak bolometric luminosities from the analytical \citetalias{yalinewich_matzner2019} modelling and from \texttt{SNEC} simulations. The last two columns also show the dimensionless prefactor $C_L$ for which the estimate in Eq. \ref{l_direct} matches the relevant luminosity. We see that the peak luminosities from the \citetalias{yalinewich_matzner2019} modelling correspond well with the \texttt{SNEC} results for most models. The most discrepant models are 11-13, for which the luminosities differ by factors of 3 to 4. 

The estimate of $C_L$ expressed from Eq. \ref{l_direct} shows a discrepancy of up to two orders of magnitude compared to $C_L$ calculated with the previous method. There is no clear trend, and larger values are obtained for models of compact structure and low density, namely models which explode off-plane (model 9-12) or with a reduced SMBH mass and scale height (model 13) or starved AGN (model 14). Excluding them and the slab model (model 8), which has discontinuous density and is challenging to simulate, leads to a factor of two spread in $C_L$ values,  $C_L \approx 7-14$ when compared against both \citetalias{yalinewich_matzner2019} models and  \texttt{SNEC} simulations (models 1,5-7,15). We note that for a spherical model such as the \texttt{SNEC} one, the spherical geometry suggests the value $C_L=4\pi \approx 12.6$ for thermalized matter, which is within this range.
 
We devote the remainder of this section to exploring predicted model lightcurves and their unique features, including comparing numerical and analytical results.

\subsection{The lightcurve shape and the effect of changing the energy and ejecta mass}

\begin{figure}
    \centering
    \includegraphics[width=8cm]{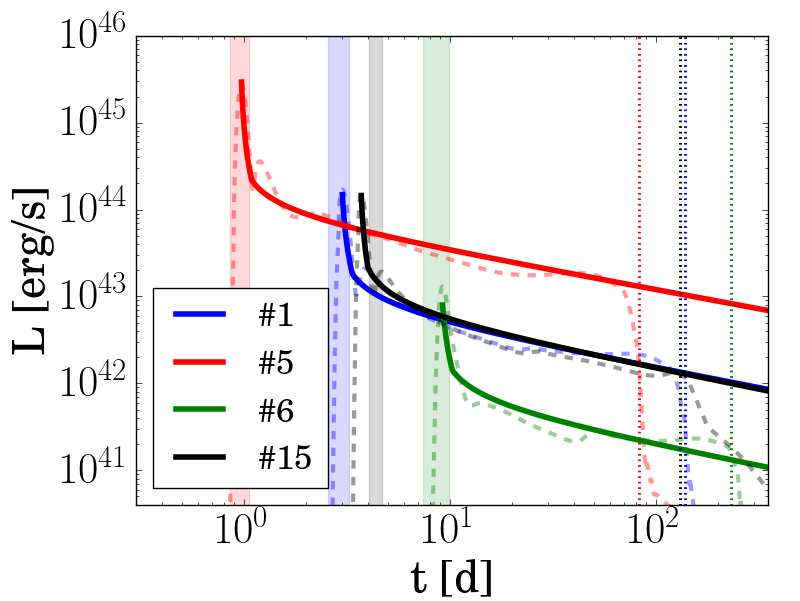}
    \caption{Probing different energies and ejecta masses. The solid lines are the analytical lightcurves where the semi-transparent dashed lines are the \texttt{SNEC} simulations. Model 1 is our canonical model (blue), model 5 has a $10$ times higher explosion energy of $E_0=10^{52}\ \rm erg$ (red), model 6 has a $10$ times lower explosion energy of $E_0=10^{50}\ \rm erg$ (green). Model 15 (black) has the same explosion energy as model 1, but with larger ejecta mass of $M_{\rm ej}=10 \msun$. The coloured transparent regions indicate the range of possible breakout times as in Fig. \ref{fig:analytica_model1}, with each transparent colour matching the correspondingly coloured model. The vertical dotted lines correspond to the end of the spherical phase for each model shown in their respective colours, also similar to Fig. \ref{fig:analytica_model1}. The effective power laws for the density for each model are $\omega_1 = 5.3$ (the same as in Fig. \ref{fig:analytica_model1}), $\omega_5=4.7$, and $\omega_6=5.75$, and $\omega_{15}=6.04$. }
    \label{fig:1_5_6}
\end{figure}

Fig. \ref{fig:1_5_6} shows the lightcurves of models 1, 5, 6, which correspond to different explosion energies ($10^{51}$, $10^{52}$ and $10^{50} {\rm erg}$, respectively) and also the lightcurve of model 15, which is identical to model 1, but with a larger ejecta mass of $M_{\rm ej}=10\ M_{\odot}$, which represents a fiducial model of core-collapse SN. We also ran a model with $10M_\odot$ ejecta and a lower energy of $10^{50}$ erg. The resulting lightcurve peak luminosity is very similar to the low energy of model 6, and is not shown here. Overall we see a good correspondence between the analytical lightcurve (solid) and the numerical results based on the \texttt{SNEC} code (dashed). We see that the larger the energy is, the faster is the breakout time, and the brighter is the peak. The dependence of the peak luminosity on explosion energy is slightly steeper than linear, $L_{\rm peak} \propto E_0 \Gamma^{(1/2+\mu\omega)\delta_{-}} \propto E_0^{1+(1/2+\mu\omega)\delta_{-}}$, which gives the value of $1.28-1.29$ for our range of density power laws $\omega_i$,  where $i={1,5,6}$. $\Gamma$ also depends on $l_{\rm eff}$, but $l_{\rm eff}$ varies only within $10\%$ between the different models. The additional corrections from $l_{\rm eff}$ slightly change the scaling to $1.23-1.24$. 

Thus, the peak luminosity reaches $\sim 3 \times 10^{45}\ \rm erg\ s^{-1}$ for model 5, while it is slightly less than $10^{43}\rm\ erg\ s^{-1}$ for model 6. The different $\omega$'s arise due to the different breakout shells in each model: the less energetic models have a lower velocity by  a factor of $E_0^{1/2}$ and hence breakout at later times. The more massive ejecta model 15 has a smaller velocity but has a longer phase of free expansion. Thus its effective power law is $\omega_{15}=6.04$, which is the largest one in our models. This is due to the fact that in this model, $l_{\rm eff}=2.67H$ is very close to the maximal value of $\omega_{\rm 15, max}=6.076$ obtained at half the photosphere $z_{\rm ph}/2H=4.93/2=2.465$. 

In the spherical phase, the dependence of the luminosity on $\Gamma$ is a very shallow power law, while the dependence on time  since the breakout is $L\propto t^{-x}$ with $x=0.44-0.47$. Note that this is steeper than the maximal slope of $-0.35$ derived by \cite{NakarSari2010} for radiation dominated material, which is obtained for $\omega=3$. This is because our extended disc model allows larger values of the effective power law index.

\subsection{Changing the SMBH mass and midplane density} \label{sec:disc_properties}

\begin{figure}
    \centering
    \includegraphics[width=8cm]{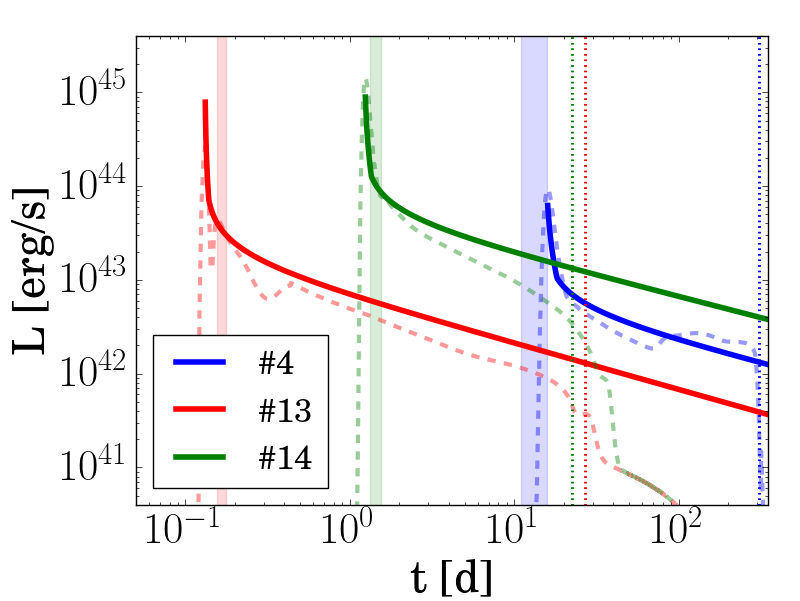}
    \caption{Probing different AGN disc locations and properties. Models 4 and 13 correspond to a $10^6 \msun$ SMBH, while model 14 corresponds to a $10^7 \msun$ SMBH. Furthermore, the explosion in model 4 is placed at $10^3 r_s$, while for models 13 and 14, the explosion is placed at $10^4 r_s$. The line style and shaded areas have the same meaning as in Fig. \ref{fig:1_5_6}. The effective power laws are $\omega_4=4.97$, $\omega_{13}=6.7$, and $\omega_{14}=4.98$.}
    \label{fig:4_13_14}
\end{figure}

Fig. \ref{fig:4_13_14} shows the lightcurves of models 4, 13, 14 where the AGN disc structure is altered. In models 4 and 13, the reduced SMBH mass of $10^6 \msun$, compared to $10^7 \msun$ for model 13, induces an increase in the midplane density, but the scale height is smaller. Model 4 is exploding at a more distant radial location of $10^4 r_s$, compared to the other two models that explode at $10^3 r_s$, thus making the swept CSM mass larger. While the analytical and \texttt{SNEC} lightcurves broadly agree for model 4, they differ more significantly for models 13 and 14. The general trend is that the \texttt{SNEC} lightcurves decrease more rapidly compared to the analytic ones. The reason may lie in the fact that models 13 and 14 have a reduced CSM interaction mass, which is slightly less than the ejecta mass.  Radiative cooling may contribute to the extra drop in the luminosity in the simulations, not accounted for in the analytical model, as discussed in sec. \ref{sub:lc}. Moreover, the peak luminosities in model 13 also differ by a factor of a few. The difference in the peak luminosities and the wiggles seen in the \texttt{SNEC} lightcurves for model 13 may be related to the reverse shock travelling back through the ejecta. Once the reverse shock reaches the origin of the spherical model, it reflects and starts travelling forward. The reverse shock arrives at the breakout region with a delay compared to the initial forward shock and is typically weaker. The reason the reverse shock is prominent in model 13 is likely because the CSM radius is small compared to other models, and the CSM mass is small enough to allow fast breakout.

\subsection{Changing vertical profile}\label{sec:omega}

\begin{figure}
    \centering
    \includegraphics[width=8cm]{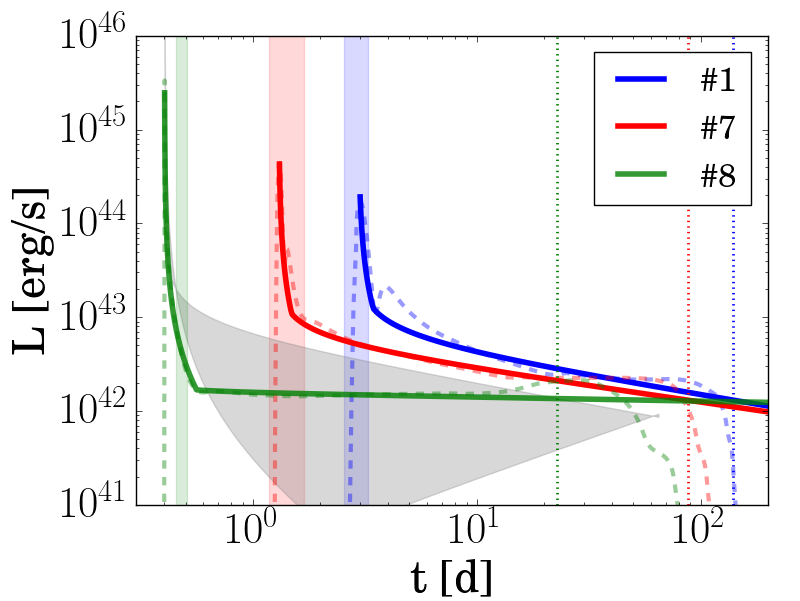}
    \caption{Probing different density profiles. The line style and shaded areas have the same meaning as in Fig. \ref{fig:1_5_6}. For models 1 and 7, we chose a different location for evaluating $\omega$, at the breakout shell, namely using  $l_{\rm edge}=z_{\rm edge} - z_{\rm bo}$, in Eq. \ref{omega_eff_rad}, where $z_{\rm edge}$ is either the physical edge of the disc (e.g. radiation-dominated profile as in model 7) or the photosphere, if the disc is formally infinite (Gaussian profile as in model 1). The spatial extent $l_{\rm eff}$ which is used to calculate $\Gamma$ is unchanged. Model 1 is our canonical model (blue), model 8 has a step profile (green), model 7 has a radiation dominated profile (red). The power-law index for model 1 is now $\omega_1^{\rm edge}=3.76$, and for model 7 $\omega_7^{\rm edge}=2.9$. For model 8, the density profile is infinite at the edge and zero elsewhere. We show the analytic fit for $\omega_8=1$ (green). The grey area indicates the possible range of lightcurve for model 8 between the two extremes: $\omega_8=0$ is the bottom limit, and $\omega_8=100$ as the top limit. The grey area extends up to the peak of the lightcurve.}
    \label{fig:1_7_8}
\end{figure}

Fig. \ref{fig:1_7_8} shows the dependence of the lightcurve on the vertical structure. We see that the radiation-dominated profile (model 7) and the uniform step profile (model 8) lead to progressively earlier and brighter breakouts. This is due to the fact that the respective edges of the different vertical profiles (and also their breakout shells) are more compact than the standard Gaussian profile in model 1. 

In Fig. \ref{fig:1_7_8}, for the calculation of $\omega$ we used  $l_{\rm edge} = z_{\rm edge} - z_{\rm bo}$ in Eq. \ref{omega_eff_rad}, where $z_{\rm edge}$ is the physical edge of the disc for model 7, and the photosphere $z_{\rm phot}$ for model 1. The spatial extent of the explosion which enters into the calculation of $\Gamma$ is unchanged, $l_{\rm eff}=z_{\rm bo} - z_1$. The resulting values for these models are $\omega_1^{\rm edge}=3.76$ for model 1, and a higher value of $\omega_7^{\rm edge}=2.9$ for model 7 (compared to $\omega_7=1.46$ in the default choice). For model 7, since $z_{\rm bo}$ is closer to the disc edge, $\omega_7^{\rm edge}$ is closer to its limiting value of $3$, which is also the power-law suitable for spherical radiative atmospheres.  For model 8, the profile is uniform and discontinuous, hence $\omega_8=0$ throughout the disc and infinite at the edge. Since $\omega_8$ is ill-defined for model 8, we express the range of possible lightcurves by shading a grey area, where the upper limit is constructed by choosing $\omega_8=100$, and the lower limit is given by $\omega_8=0$. For comparison, we also plot the analytic fit with $\omega_8=1$, which gives a roughly flat lightcurve, similar to the \texttt{SNEC} lightcurve (though we caution that a numerical treatment will also struggle with a discontinuous density profile, and the \texttt{SNEC} models are likely not robust in this case).


For model 1, this fit is comparable to the one with the default choice (as in Figs. \ref{fig:analytica_model1} and \ref{fig:1_5_6}). This is also a representative case of all the gas dominated Gaussian profiles, namely that they are not very sensitive to $\omega$. For radiation dominated profiles, however, the latter  choice of $\omega_7^{\rm edge}$ gives a much better match to the \texttt{SNEC} results. We speculate that the reason may be due to the fact that radiative dominated and slab profiles are more compact and have an actual edge.

\subsection{Changing vertical location}

\begin{figure}
    \centering
    \includegraphics[width=8cm]{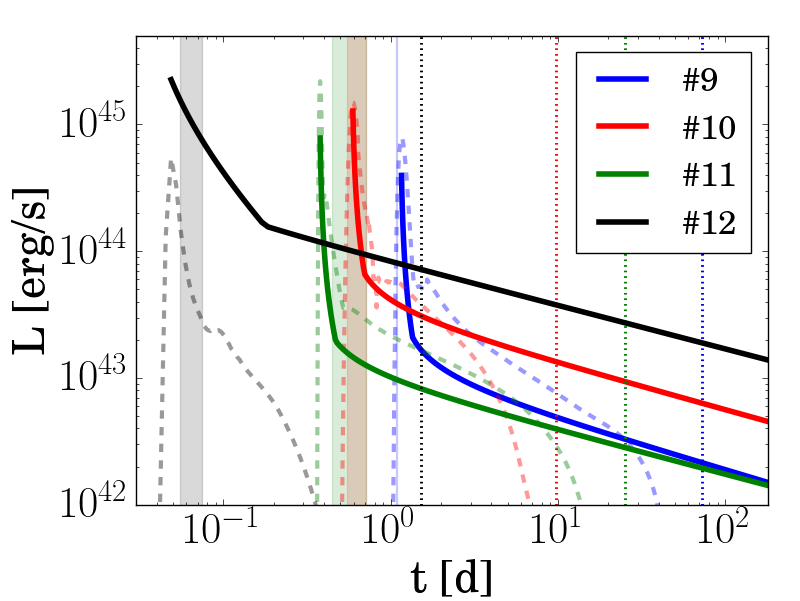}
    \caption{Changing the vertical location of the explosion. Gaussian profiles are for $z_0=1H$ (blue, model 9) and $z_0=2H$ (red, model 10). Radiative profiles are for $z_0=1H$ (green, model 11) and $z_0=2H$ (black, model 12). The line style and shaded areas have the same meaning as in Fig. \ref{fig:1_5_6}. We use the $\omega^{\rm edge}$ choice as in fig. \ref{fig:1_7_8}. The effective power laws are $\omega^{\rm edge}_9=3.6$, $\omega^{\rm edge}_{10}=3.17$, $\omega^{\rm edge}_{11}=2.93$, $\omega^{\rm edge}_{12}=2.87$.}
    \label{fig:9_to_12}
\end{figure}

Fig. \ref{fig:9_to_12} shows the lightcurves of the off-plane explosion models 9-12. Among the Gaussian density profiles, the further out explosion (model 10) at $z_0=2H$ explodes earlier and is brighter, as expected. On the other hand, for radiative density profiles, the further explosion (model 12) explodes earlier, but is dimmer than the $z=1H$ case (model 11). The reason lies in the very low CSM mass for model 12 ($M_{\rm CSM} < 10^{-3}M_\odot$). 

The predicted breakout times are compatible with the \texttt{SNEC} models, although for the $z_0=2~H$ cases, the breakout occurs slightly earlier than expected. The reason may lie again in the low CSM mass, which limits the thermalization of the CSM material.  This  could account for the large discrepancy between the analytic and numeric lightcurves in model 12.

\subsection{Comparison between the codes} \label{comparison}

\begin{figure*}
    \centering
    \includegraphics[width=17cm]{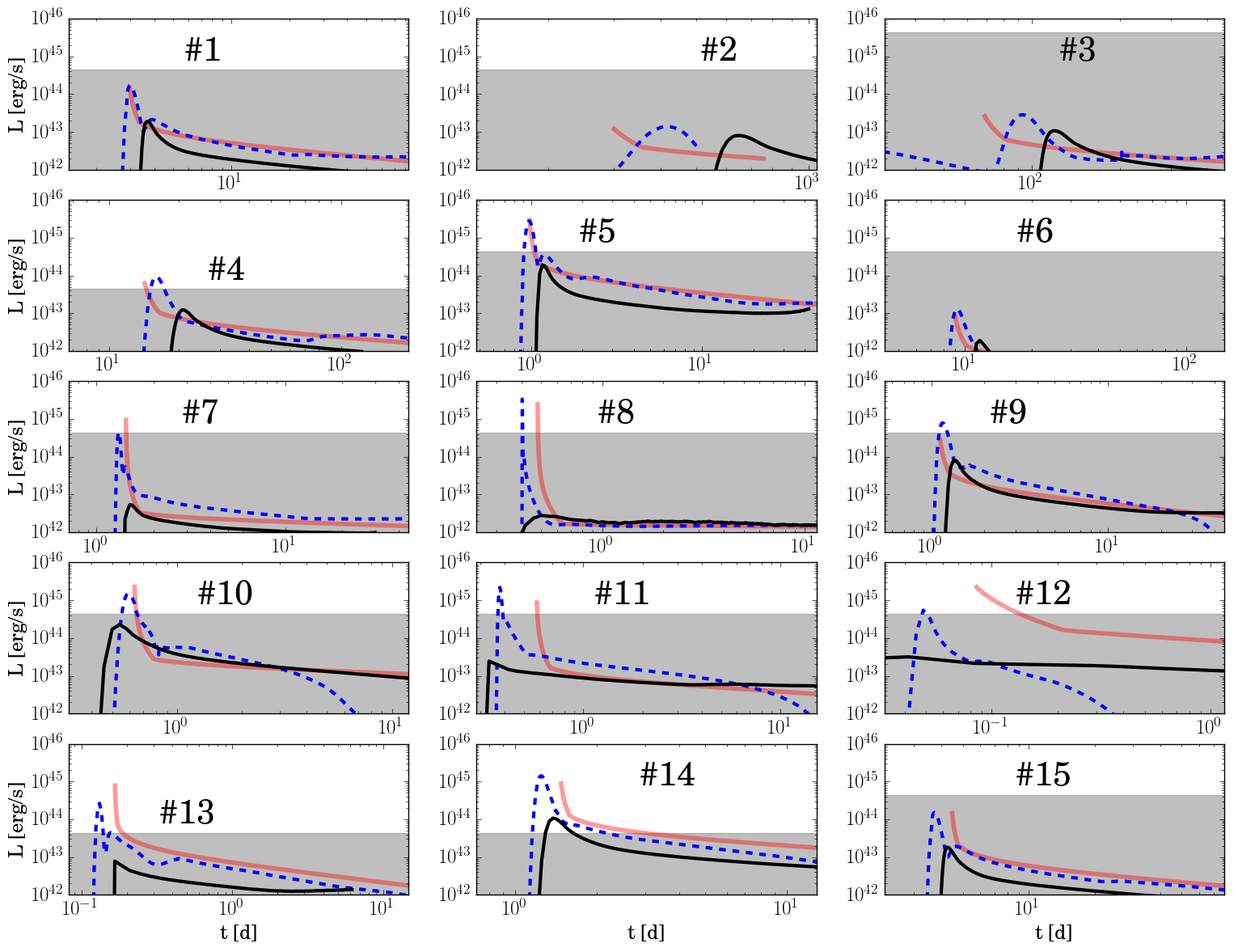}
    \caption{ Comparison between the \texttt{SNEC} code (dashed blue) and the \texttt{HORMONE} code (solid black) for all sampled models. The red curve is the analytic fit, but starting with the peak at the average breakout time of the two velocity models calculated in Appendix \ref{appendix-breakout}, $t_{\rm bo}= (t_{\rm bo}|_{\rm pw} + t_{\rm bo}|_{\rm MM99})/2$. The grey area is where the AGN luminosity can obscure the transient. The maximal AGN luminosity is set according to Eq. \ref{l_agn}. The model number appears in bold black for each panel. Note that the ordinate is the same, but the abscissa is different for each panel.}
    \label{fig:comparison}
\end{figure*}

Fig. \ref{fig:comparison} shows the lightcurves of different models and compares predictions from the Lagrangian spherically-symmetric radiative hydrodynamics \texttt{SNEC} code (dashed blue) to those from the Eulerian 2D hydrodynamics \texttt{HORMONE} code (solid black). The goal is to understand how important are the effects of the more realistic 2D disc geometry, including its structure evolution and morphology, since our analytic and \texttt{SNEC} models assume spherical symmetry. The \texttt{HORMONE} simulations do not include radiative transfer and thus radiative cooling and the transfer of energy from the inner parts of the CSM outwards are not properly modelled. However, this only becomes important in the later phases and the dynamics up to and close to shock breakout should be mostly adiabatic. 
The vertical shock propagation in the 2D \texttt{HORMONE} simulations closely resembles that of the 1D \texttt{SNEC} simulations as shown in Fig.~\ref{fig:velocity_models}. 

The lightcurves for the \texttt{HORMONE} models in Fig.~\ref{fig:comparison} are all computed for a face-on viewing angle. This is expected to be the angle where the shock breakout is observed with the brightest luminosity. Unlike the spherically symmetric models, the shock breakout in the 2D models starts from a small patch and spreads out radially due to the different shock propagation times along each polar angle. Furthermore, the effective CSM mass the shock has to interact with before breaking out also increases with the polar angle, decreasing the velocity at the breakout shell and making the luminosity at corresponding annuli lower. These differences may likely explain the lower luminosities reached in \texttt{HORMONE} models in Fig.~\ref{fig:comparison}. The peak is also smeared out over a slightly longer duration because of the delay in shock breakout at lower latitude angles. 

For the off-plane explosions, the difference in interacting mass along each polar angle is smaller compared to the mid-plane explosion models. Therefore, the opening angle of the shock breakout region is wider, leading to more similar peak luminosities between 2D and spherical models (Fig.~\ref{fig:comparison} panels 9, 10). 
 
\subsection{Summary of the results}
In summary, the typical lightcurve of a SN explosion will peak at luminosities around $10^{44} - 10^{45} \rm erg\ s^{-1}$. The earliest peak will usually be the brightest one. Radiation-dominated or slab profiles will result in more compact and less massive vertical layers and could produce stronger observable events when compared to gas pressure-dominated (Gaussian) profiles. The type of the SN (or the ejecta mass) will play a relatively minor role in the observational signature, with the main parameter being the explosion energy. This is the case so long as most of the kinetic outflow encounters enough CSM material to reprocess the kinetic energy into outgoing radiation.

The strongest peak will be reached if the CSM mass is comparable to or smaller than the ejecta mass (models 8, 12, 13, 14), although models with CSM mass exceeding the ejecta mass by a factor of ten are also observable. Models with too much CSM mass ($\sim 100 M_{\rm ej}$) will be choked and will not be observable. For these reasons, off-plane explosions, explosions in the inner ($\sim 10^3 r_s$) regions of AGN discs and explosions in the discs of starved AGNs will lead to the brightest transients.

\section{Discussion} \label{sec:discussion}
\begin{figure*}
    \centering
    \includegraphics[width=0.3\linewidth]{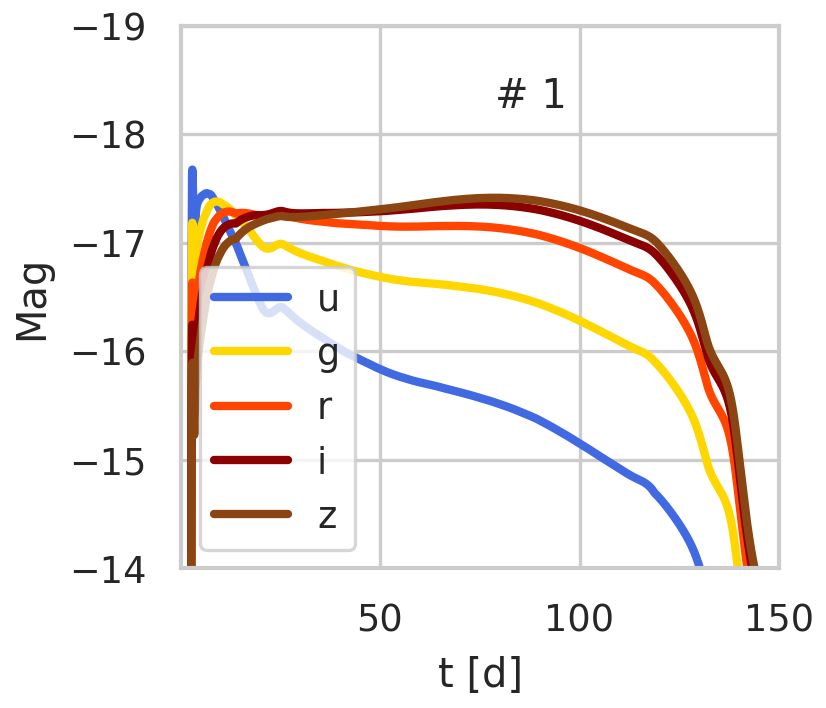}    \includegraphics[width=0.3\linewidth]{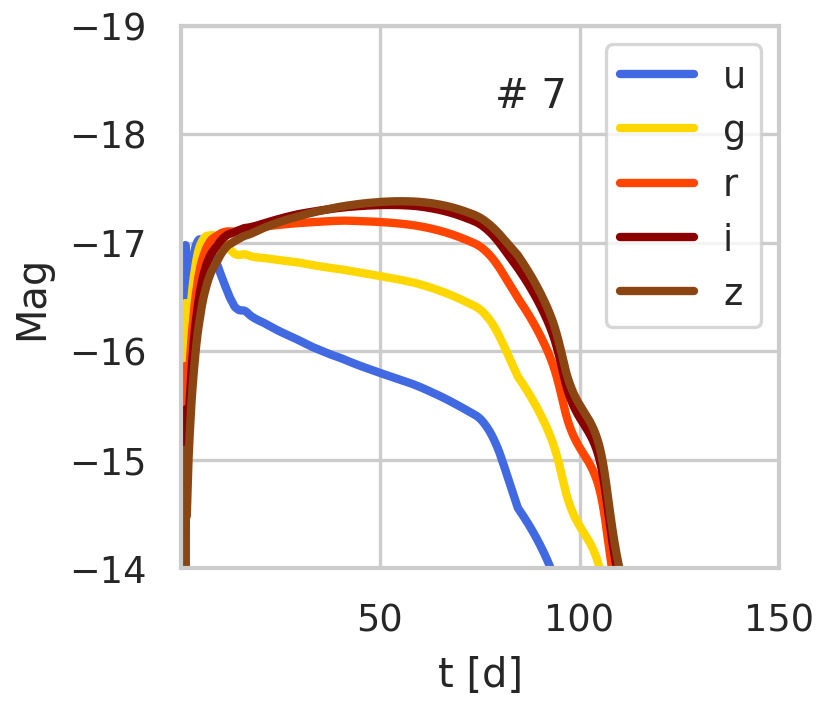}  
    \includegraphics[width=0.3\linewidth]{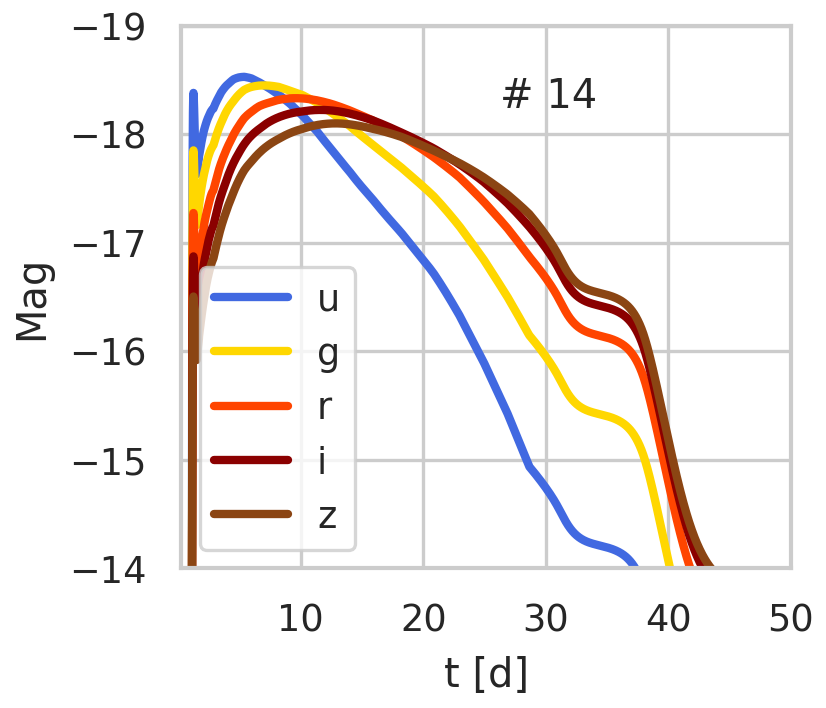}
    \caption{SNEC-based ugriz lightcurves for models 1, 7 and 14. The models represent the CSM interaction signature for a type Ia supernova in a typical AGN around a $10^7\,{\rm M}_\odot$ black hole for a pressure-dominated (model 1) and radiation-dominated (model 7) atmospheres, and the same supernova in a starved AGN (model 14). The initial blue/UV half-day-long spike due to shock breakout (which produces the brightest signal in extreme UV / soft X-rays) is followed by the week-long blue/white peak, followed by an extended red/IR emission from the deposited shock energy reprocessed by the CSM. The shape and fall-off times of the lightcurves after the main peak may be inferred from the bolometric lightcurves.}
    \label{fig:ugrizLCs}
\end{figure*}

\subsection{Observational picture}

Although the SN explosions in AGN disks are typically more luminous than identical explosion in the absence of circumstellar material, they still need to outshine the luminous AGN environment. \citet{Hubeny2001} studied the spectra of AGN accretion discs and found that for a $10^7 \msun$ disc with a viscosity parameter $\alpha=0.01$ and luminosity
\begin{equation}
    L=0.3L_{\rm Edd}= 4.4 \cdot 10^{44} \left( \frac{M_{\bullet}}{10^7 \msun} \right) \rm erg\ s^{-1} \label{l_agn},
\end{equation}
  the spectrum peaks at $\nu F_\nu \approx 10^{44}\ \rm erg\  s^{-1}$ in the extreme UV / soft X-ray band around $0.1\ \rm keV$. In Eq. \ref{l_agn}, the Eddington luminosity $L_{\rm Edd}$ assumes the electron scattering opacity.

\textbf{Bolometric luminosity:} For our suite of models, models 1, 6, 9 and 15 peak around $\sim 10^{44} \rm erg\ s^{-1}$ or lower and will likely be obscured by the AGN luminosity in the absence of clear features that make it possible to distinguish AGN spectra from the explosion spectra.  Model 4 also peaks around $\sim 10^{44}\ \rm erg\ s^{-1}$, but the host AGN is less massive. Thus it is a factor of $\sim 10$ more luminous than the AGN background. Models 5,  8, 10, 11 and 14 peak with luminosities $L\gtrsim 10^{45}\ \rm erg\ s^{-1}$ and are thus most likely to be observed. Models 13 and 14 are especially promising since they are located in a low-density environment, either in a less massive galaxy or in a starved AGN environment, and could be more luminous than their respective background by a factor of $\gtrsim 100$. We note that this is a conservative estimate, since the observed luminosity can be much lower ($\sim 0.01L_{\rm Edd}$, \citealp{fabian2009}), which will make all models besides 3,4, and 6 observable. 

In summary, the events with the best chances to be observed will be either in a radiation-dominated disc, and/or in a low-density environment. There are three possibilities to achieve this among the set of models we considered:  i) in a low mass galaxy ($M\approx 10^6 \msun$, ii) in a starved AGN of reduced density, or iii) explosions away from the midplane. 

\textbf{Multiband lightcurves:} We show the typical multi-band lightcurves in Fig.~\ref{fig:ugrizLCs}. For all the models, the first breakout will produce a short UV/blue transient, followed by a red/IR tail. As mentioned above, the initial breakout peaks in extreme UV / soft X-rays, so these lightcurves understate the peak bolometric luminosity. The emission is mostly optical after the early breakout peak, which may help to separate these events from the AGN background, which peaks in UV/X-rays. Furthermore, the models producing sharp peaks, e.g. models 8, 12, 14, reach temperatures high enough to give rise to X-ray flares. In this case, we expect approximately hour-long flares reaching up to $10^{43} \rm erg/s$ in the $0.3$~--~$10\,\rm keV$ band. The exceptions to this picture are models 2 and 3, in which the shock stalls due to the large CSM mass. In this case, the thermalised energy is emitted over hundreds of days in red/IR bands.

The presence of $^{56}$Ni in real transients will lead to additional energy deposition on a month timescale. However, since the energy is deposited mostly in the ejecta material located behind the CSM, the luminosity contribution from $^{56}$Ni will be delayed by the photon diffusion time and eventually emitted over several months or longer in red/IR bands. Since the total energy yield in $^{56}$Ni is comparable to that of the regular nuclear transients in the field and since this energy is emitted over longer times, we find that the contribution from radiative decay typically does not exceed $10^{43} \rm erg/s$. The AGN will typically outshine such red/IR contributions. Conversely, in models 12-14 with low CSM mass, the CSM lightcurve after the peak becomes quickly dominated by the contribution from the nuclear decay. In this case, the late lightcurve resembles that of isolated supernovae.

The explosion and its ejecta are expected to be non-relativistic. It is possible that some fraction of the electrons, especially where the temperature is large enough, will not be in thermal equilibrium and will be accelerated to relativistic velocities producing $\gamma$-rays, either in GRBs \citep{GRB_in_AGN,perna20} or in hyper-Eddington accretion-induced Bondi explosions \citep{wang2021}, where due to
inefficient energy transport the material is heated to very large temperatures, and relativistic shocks make a large cavity in the AGN disc. While the GRB radiation is a prompt emission, the radiation from Bondi explosions will be visible only in the broad-line region, $\sim 1\ \rm yr$ after the explosion. For GRBs, following the cocoon breakout, non relativistic ejecta also break out at later times. We find similar peak luminosities to the estimates of \cite{GRB_in_AGN}. 

\subsection{Rates of SNe in AGN discs} 

\begin{figure*}
    \centering
    \includegraphics[width=17cm]{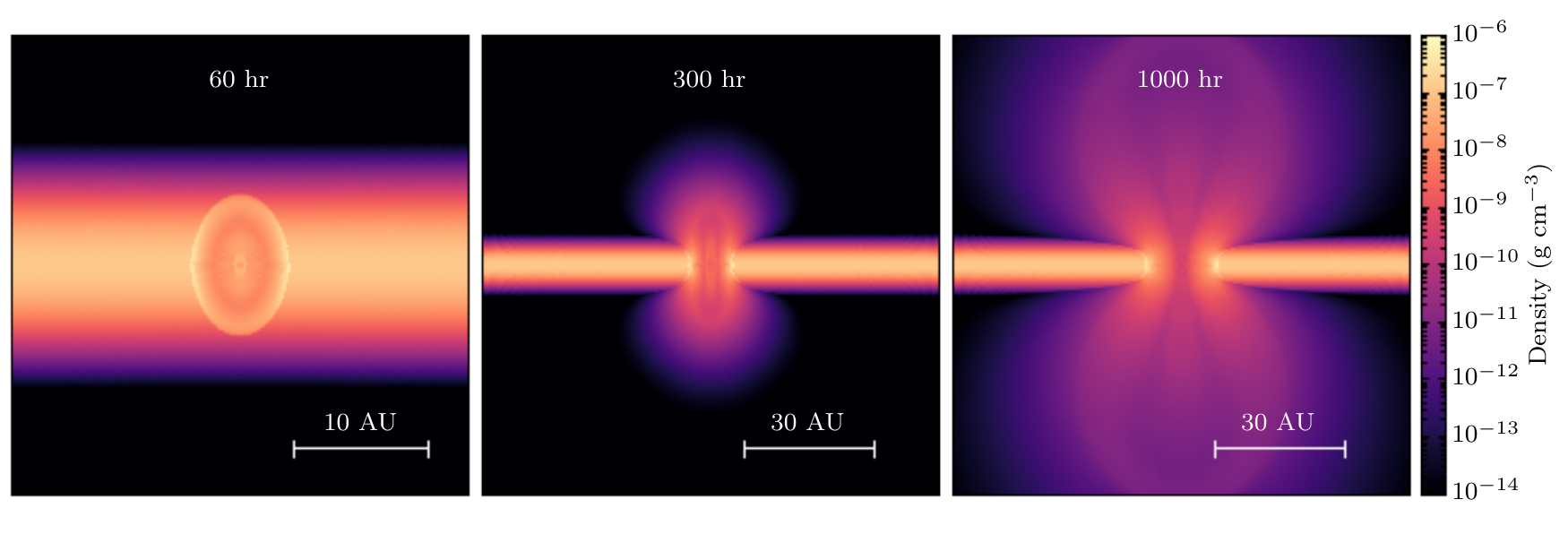}
    \caption{Density snapshots from the 2D simulations for model 1. Each panel shows a different time snapshot. Note that the left panel has a smaller box size.}
    \label{fig:model1_2d}
\end{figure*}

\begin{figure*}
    \centering
    \includegraphics[width=17cm]{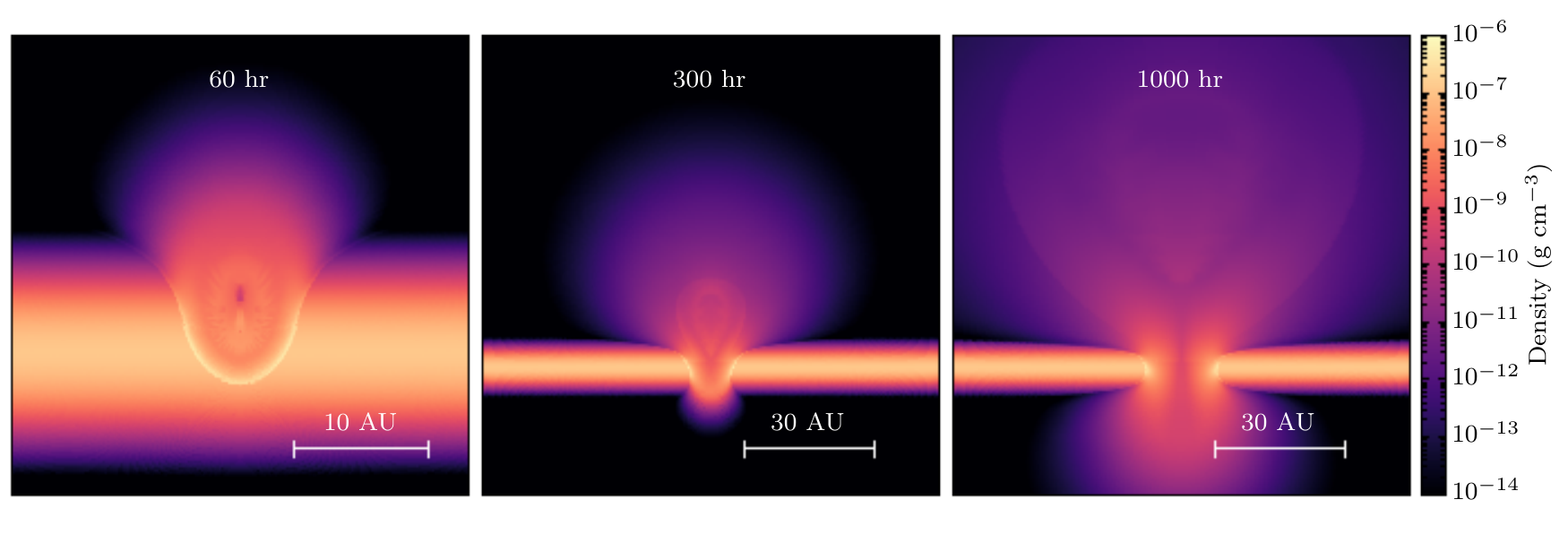}
    \caption{Same as Fig.~\ref{fig:model1_2d} but for model 9. }
    \label{fig:model9_2d}
\end{figure*}

The radiative efficiency of an accreting AGN is $\eta = L/(\dot{M}c^2)<1$.  If the SMBH is accreting at a fraction $L=fL_{\rm Edd}$ of the Eddington luminosity, and the same fraction $f$ of the mass accretion rate, the SMBH mass doubling time is 
\begin{equation}
    t_{\rm doubling} = \frac{M_\bullet}{\dot{M}} =\frac{\eta}{f} \frac{\kappa c}{4\pi G} = 128 \left( \frac{\eta}{0.1} \right) \left( \frac{f}{0.3} \right)^{-1} \rm{Myr}.
\end{equation}
For a disc lifetime of $t_{\rm AGN}=10^8\ \rm yr$, the mass that flows into the disc is $M_\bullet t_{\rm AGN}/t_{\rm doubling} \approx M_\bullet$.

The type Ia SN rate is 1 per few hundred solar masses of star formation in the field \citep{Maoz2017}. Assuming a similar but slightly higher efficiency for AGN discs due to an enhanced binary merger rate (e.g., one SN event per $100 \msun$), and further assuming a fraction $f_1$ is converted into stars, we would have $f_1 10^5$ SNe for a SMBH mass of $10^7 M_\odot$ over an AGN lifetime $t_{\rm AGN}=10^8\ \rm yr$. Thus, the rate per AGN is $f_1 10^{-3}\ \rm yr^{-1}$  per AGN disc. Roughly 1\% galaxies in the local Universe have an AGN disc, so the AGN density is $10^{-4}\ \rm Mpc^{-3}$. Thus, we expect a total of $\sim f_1 100$ AGN SNe per $\rm Gpc^3$ per year. A similar estimate for the rate of core-collapse SNe leads to a slightly higher but otherwise comparable rate of events. This rate is also slightly larger than the the upper limit of the binary black hole merger rate in AGN discs, $\mathcal{R}=0.002-18\ \rm Gpc^{-3}\ yr^{-1}$, which had been recently estimated by \cite{2grobner2020}.

A proper rate estimate should include an integral over the mass function of the AGN in the local universe. Moreover, indirect observations of the AGN phase duration ($10^6-10^9\ \rm yr$), and hence the duty cycle ($10^{-1}-10^{-3}$), could vary by around one order of magnitude \citep{martini2001, marconi2004}. Furthermore, AGNs may not be accreting steadily but rather experience events of rapid accretion and quiescence every $10^5$ years \citep{agn_flicker} which could affect the star formation rate.

\subsection{Observable event rates}

What fraction of the SNe will be observable? In order to address this question, we need to discuss the spatial distribution of stars in AGN discs, the mass function of the formed stars and its dependence on accretion. 

Star formation in AGN discs predominantly occurs in the outer regions of the disc at low temperatures, where the disc is gravitationally unstable. The star formation efficiency is especially large when the midplane temperature is between  $ 10^3 - 10^4\ \rm K$, and the opacity is very low since dust is still vaporized, but the material is mostly neutral, so electron scattering is low. 

\textbf{Migration:} The interaction of the stars with the massive gaseous AGN discs can induce torques on them and cause the stars to migrate \citep[e.g.][and references therein]{artymovicz1993,Mck+12,bellovary2016}. Migration can be stopped if the torque changes sign. Extensive work had been done on migration in protoplanetary discs \citep{goldreich1980, tanaka2002, paardekooper2006, paardekooper2010}, and migration traps were numerically reproduced \citep{lyra2010}. \cite{bellovary2016} found that the outermost migration traps for AGN discs lie at around $\sim 300 r_g$. If most stars end up in this region, the binary mergers and SN explosions there will be both more frequent and more observable.

However, AGN discs are not necessarily as well behaved as protoplanetary discs. The immense radiation and mass make them prone to flaring and instabilities. It is uncertain whether the disc structure of \citetalias{sg03} and \citet{tqm05} is in a steady state, and the fitting formulae for the torque as in \citet{paardekooper2010} may not be applicable. Moreover, \citet{bellovary2016} used an adiabatic index of $\gamma=5/3$. Radiation-dominated discs with $\gamma=4/3$ will result in a different location for the migration traps, which may be shifted outwards.

Another issue is the timescale for the migration to take place. For type I migration, in which viscous torques are unable to form a gap in the disc, the migration time is \citep{tanaka2002}
\begin{equation}
    \tau_I \sim \frac{M_{\bullet}^2}{\msun \Sigma(r)r^2} \left(\frac{H}{r}\right)^2 \Omega^{-1} = 7 \left(\frac{M_{\bullet}}{10^7 \msun} \right)^2 \left(\frac{r}{10^4r_s} \right)^2\ \rm Myr  \label{t_typeI}
\end{equation}
Stars that formed beyond $10^5 r_s$ or in more massive discs would require longer than the lifetime of the disc to migrate, while less massive discs enable faster, more efficient migration. In some very low mass discs, a gap may be carved, and the migration timescale will be of type II, which will not depend on the mass of the disc (or the SMBH mass). Stars may still be captured from the nuclear star cluster onto the AGN disc.

Finally, the dynamical friction experienced by the stars could be significantly different from these simple estimates if the accretion onto the stars produces strong outflows \citep{gruzinov2020}, which is almost inevitable for the AGN environment where the accretion rates can be well above the Eddington limit (also see below). 

The spatial distribution of the progenitors of SN explosions will depend on the migration history. Since the shocks from explosions at radial distances beyond $r\gtrsim 10^3r_s$ will be choked, the number of observed progenitors in the disc will be $N_3=N(r\lesssim 10^3\ r_s)$. If migration is effective, most of the progenitors may be close to the migration traps, yielding a fraction of observed events $f_2=N_3/N_{\rm tot} \sim 1$, where $N_{\rm tot}$ is the total number of the progenitors. If no migration occurs, then assuming an underlying log-uniform distribution between $10^2-10^5\ r_s$, only $f_2 \sim 0.3$ of the events will be observable. If most stars are either stuck in the outer regions or migrate outward, then $f_2\ll 1$ could be rather low. 

\textbf{Accretion:} Accretion could be important for changing the initial mass function of stars in AGN discs and affecting the type Ia / core-collapse SN branching ratios. However, neither the details of the expected super-Eddington accretion nor the initial mass function are certain. \cite{artymovicz1993} first studied the growth and accretion of stars in AGN discs. More recently, \citet{davies2020, cantiello1} suggested that significant growth can occur. \citet{Lei+13} also had a similar discussion, with possible implications for depleting the gas through accretion. However, a recent follow-up study showed that rotation and radiative feedback could limit the final mass to be around $10\ \rm M_{\odot}$ \citep{cantiello2}, especially for the innermost regions, which we mainly consider. Moreover, the extremely optically thick environment may block the energy release via radiation and inhibit the accretion rate. A similar situation could lead to "Bondi explosions" \citep{wang2021}.

Accretion onto white dwarfs might, under appropriate conditions, allow for their growth to close to the Chandrasekhar limit and eventual explosion as type Ia SNe \cite{Ost+83}, thus increasing the rate for such events. However, here too, the exact accretion efficiency is difficult to assess.

The frequency of off-plane explosions is also quite uncertain, but probably low. Gas dynamical friction tends to damp the orbital inclination much faster than the migration timescale \citep{rein2012,grishin2015} if the orbital inclination with respect to the disc midplane is not too large. Highly inclined orbits $(i\gg H/r)$ may take much longer to dissipate into the plane of the disc, but they spend most of their time outside of the disc, and are therefore likely to explode in relative isolation without significant CSM interaction. Accretion feedback may change this picture \citep{gruzinov2020}, while close encounters with neighbouring stars may excite the progenitors' orbits.

To summarize, the upper limit for the rates of detectable events is $\mathcal{R} \lesssim f_1 f_2 100 \ \rm yr\ Gpc^{-3}$, where $f_1$ is the efficiency of forming AGN gas into stars, and $f_2$ is the fraction of observable events. The fraction of core-collapse SNe is uncertain and could constrain the efficiency of accretion or the underlying mass function.

Remaining agnostic to the efficiency of accretion and its final fate, we comment that for a Salpeter mass function $dN/dm\propto m^{-2.35}$, around $1\%$ of stars are born massive enough to produce core-collapse SN explosions, while for a top-heavy mass function $dN/dm \propto m^{-0.45}$ \citep{bartko2010}, the fraction of massive stars increases to $\sim\ 10\%$, thus increasing the rates by an additional factor of ten. 

\subsection{Disc feedback}

The 2D hydrodynamical simulations allow us to explore the morphology of the ejecta and the impact on the local disc morphology around the explosion site. Fig.~\ref{fig:model1_2d} shows density snapshots of model 1. We see that at the early time of  $60\ \rm hr$, shortly before shock breakout (left panel), the blast wave is still rather spherical, expanding within the disc. The shock front is gradually deformed as the in-plane directions are pinched, and the shock front propagates faster in the vertical directions, making the blast wave take a prolate form. Later, the shock breaks out from the disc surface, and a local hole is carved in the disc. The ejecta flow out almost spherically from the location of shock breakout. 

 The overall features are qualitatively similar to recent 3D numerical simulations of an SN explosion in a more distant radial location \citep[$M_\bullet=10^8~\msun, r=2\cdot10^4~r_s$;][]{morenchel2021}. Due to the larger spatial extent and higher interaction mass, the more distant event will not be observable. However, SN explosions may contribute to the global viscous evolution of the AGN disc. The 2D simulations we present do not take into account the shearing motions of the disc. In reality, the hole excavated by the SN will be elongated in the disc rotation direction \citep{morenchel2021}.

Fig.~\ref{fig:model9_2d} shows the density snapshots of the off-plane explosion, model 9. We see that the shock breaks out earlier from the top edge, while the downward shock propagates longer as it climbs up the density gradient. This downwards CSM interaction creates a strong upwards reverse shock, which quickly flows out through the upper hole. The downwards shock also eventually breaks out from the other side of the disc. The breakout from this side of the disc is much weaker than the upper side, and the luminosity should be much dimmer.

\subsection{AGN flaring and variability}
The significant variability observed for AGNs over a wide range of timescales gives rise to difficulties in identifying nuclear transients and separating such transients from various forms of inherent AGN variability. 

As discussed above, the observational signatures of SNe in AGN discs resulting from our models suggest that the SN transients will often be challenging to observe above the background luminosity of the AGN. However, in some cases, these AGN SNe can surpass the AGN background during their rapid rise to a peak. The timescale for such initial high-luminosity flaring is of the order of hours up to 1-2 days at most (see sec. \ref{sec:results}). Identifying such short-term transients requires high-cadence transient surveys. Moreover, many transient surveys do not focus on AGN nuclear regions where the AGN variability challenges identification. To the best of our knowledge, the only high-cadence sky surveys of AGNs were done using the Kepler mission. Interestingly, these surveys did identify one fast, days-timescale flare event KIC 1606852 \citep{Smi+18}, which could potentially be related to the AGN SNe we discuss. However, it might be too long or be the result of other explosive processes \citep{Smi+18}. The timescale of other nuclear transients, such as candidate tidal disruption events and changing-look AGNs, are far longer than the peak timescale of AGN SNe and are not likely to be related. Future high-cadence transient surveys may be able to identify AGN SNe.

\section{Summary} \label{sec:summary}

The AGN disc environment around SMBHs has dense stellar populations, many of which are expected to be embedded in the disc. As stars end their lives in supernova explosions, the dense AGN environment makes a fertile ground for unique transient events. Understanding such explosions and their properties opens a window to the physics of AGN discs and their interactions with the supernova progenitors.

Motivated by superluminous explosions in circumstellar material (CSM), we have developed an analytical model for the expected time of breakout and subsequent lightcurve, depending on the disc and progenitor properties.  We validated this model with extensive numerical simulations. We found that the typical peak luminosity for such events may reach a few times $10^{45}\ \rm erg\  s^{-1}$. The most energetic events are also the quickest to break out, ranging from hours to several days, until they become too faint to be detected. 

The most luminous explosions are generally found either in radiation-dominated discs, at larger explosion energies, or in regions with reduced density, such as off-plane explosions, low mass SMBH, or starved AGNs with reduced density. The latter two (i.e. low SMBH mass and starved AGN discs) also have better chances of being observed due to reduced background AGN luminosity. The initial breakout events should be dominated by the blue bands, while at later times, the lightcurves will be dominated by the red and infrared bands. The upper limit for the event rate is $\mathcal{R} \lesssim 100\ \rm yr\  Gpc^{-3}$, where optimal star formation and observational conditions are assumed. The actual rate could be lower by orders of magnitude. 

In concise form, our study may be summarised as follows: {\it We exploded stars in an AGN.  The photons can't get out, the outflow can. The gas converts shocks into brighter SN. We may see a few hundred, if we frequently scan.}

\section*{Data availability}
The simulations underlying this article will be shared on reasonable request to the corresponding author.

\section*{Acknowledgements}
We thank the referee, Pablo Fabi{\'a}n Vel{\'a}zquez, for valuable comments on the manuscript. We thank Almog Yalinewich, Ari Laor and Yossef Zenati for stimulating discussions. EG and HBP acknowledge support for this project from the European Union's Horizon 2020 research and innovation program under grant agreement No 865932-ERC-SNeX. IM is a recipient of the Australian Research Council Future Fellowship FT190100574. 





\bibliographystyle{mnras}



\appendix

\section{Vertical profile for radiation-dominated pressure} \label{vertical profile}

Consider the radiation pressure   $P=aT^{4}/3=K\rho^{4/3}$, where $a=4\sigma_{\rm SB}/c$, $\sigma_{\rm SB}$ is the Stefan-Boltzmann constant and $c$ is the speed of light.  We assume a polytropic equation of state where $P=K\rho^{\gamma}$. The sound speed is given by $c_{s}=\sqrt{dP/d\rho}=\sqrt{\gamma K\rho^{\gamma-1}}=\sqrt{\gamma P/\rho}$. The vertical hydrostatic equilibrium equation is 
\begin{equation}
\frac{dP}{dh}=-\rho g_z=K\gamma\rho^{\gamma-1}\frac{d\rho}{dh}
\end{equation}
where $g_z=GM_{\rm SMBH}h/(r^2+h^2)^{3/2}$ is the $z$ component of the tidal gravitational acceleration. Since the disc is thin ($h\ll R$), we can approximate it as $g_z \approx \Omega^2 h$ where $\Omega^2=GM_{\rm SMBH}/r^3$ is the Keplerian frequency. We have the ordinary differential equation
\begin{equation}
\rho^{\gamma-2}d\rho=-\frac{\Omega^{2}}{\gamma K}. 
\end{equation}
Integrating, we have 
\begin{equation}
\frac{\rho^{\gamma-1}}{\gamma-1}=-\frac{\Omega^{2}}{2\gamma K}z^{2}+C.
\end{equation}
To determine $C$ we use the initial condition at $h=0$ 
\begin{equation}
\rho^{\gamma-1}=(\gamma-1)C=\rho_{0}^{\gamma-1} \implies C=\rho_0^{\gamma-1}/(\gamma-1).
\end{equation}
Finally, the vertical profile is 
\begin{equation}
\rho(h)=\rho_{0}\left[1-(\gamma-1)\frac{\Omega^{2}}{2\gamma K\rho_{0}^{\gamma-1}}h^{2}\right]^{1/(\gamma-1)}.
\end{equation}
Note that the scale height $H$ is defined as
\begin{equation}
H^{-2}=\frac{\Omega^2}{c_s^2}=\frac{\Omega^{2}}{\gamma K\rho_{0}^{\gamma-1}},
\end{equation}
which leads to the final expression
\begin{equation}
\rho_{\rm rad} (h)=\rho_{0}\left[1-(\gamma-1)\frac{h^{2}}{2H^{2}}\right]^{1/(\gamma-1)}.
\end{equation}
Note that the radiation pressure-dominated profile is steeper than the Gaussian profile $\rho_{\rm gas} \propto \exp(-h^2/2H^2)$ for gas dominated pressure $P\propto \rho$, and goes to zero at $h/H=\sqrt{2/(\gamma-1)}$,  or $h/H=\sqrt{6}\approx2.45$. 

\section{Breakout times in AGN discs} \label{appendix-breakout}

\textbf{Gaussian density profile:}
For the piecewise velocity (Eq. \ref{vpi}) in the Gaussian density profile (Eq. \ref{eq:rho_gas}) the integral in Eq. \ref{eq:t_bo} can be split into three integrals. The first one over the free expansion phase is trivial, while the other two involve integrals of the form $\int x^{3/2}e^{-\alpha x^2/2}dx = - 2^{1/4}\Gamma(5/4, \alpha x^2/2)/\alpha^{5/4}$, where  $\Gamma(s,x)=\int_x^\infty y^{s-1}e^{-y} dy$ is the upper incomplete Gamma function. We get

\begin{align}
    \left.\frac{t_{\rm bo}}{H/v_0}\right|_{\rm pw}=&\zeta_1 + \frac{2^{1/4} \mathcal{A}_1}{\zeta_{1}^{3/2}\mu'^{5/4}}\left[\Gamma\left(\frac{5}{4},\frac{\mu'\zeta_{1}^{2}}{2}\right)-\Gamma\left(\frac{5}{4},\frac{\mu'\zeta_{2}^{2}}{2}\right)\right]  \nonumber \\
    +&\frac{2^{1/4} \mathcal{A}_2}{\zeta_{1}^{3/2}\mu{}^{5/4}}\left[\Gamma\left(\frac{5}{4},\frac{\mu\zeta_{2}^{2}}{2}\right)-\Gamma\left(\frac{5}{4},\frac{\mu\zeta_{{\rm bo}}^{2}}{2}\right)\right]
    \label{t_pw}
\end{align}

where $\zeta_i,\zeta_{\rm bo}=z_i/H, z_{\rm bo}/H$ are dimensionless length scales,   $\mathcal{A}_1=(\rho_0/\rho_1)^{\mu'}$ and $\mathcal{A}_2 = (\rho_2/\rho_1)^{\mu'}(\rho_0/\rho_1)^\mu$ are coefficients that depend on the density at each location, which comes from the continuity requiremtns of the piecewise velocity in Eq. (\ref{vpi}).  For the parameters of model 1, we get $t_{\rm bo}=3.17\ \rm d$.

For the \citetalias{MM99} velocity model (Eq. \ref{vMM99}) in the Gaussian profile, the breakout time is given by 
\begin{equation}
    \left. t_{\rm bo} \right|_{\rm MM99}= \frac{H}{v_{0}}\intop_{0}^{\zeta_{{\rm bo}}}\left(1+\frac{\rho_{0}H^{3}}{M_{{\rm ej}}}\zeta^{3}\right)^{1/2}e^{-\mu\zeta^{2}/2}d\zeta \label{t_mm99}.
\end{equation}
This is not an analytic expression, but can be evaluated numerically. For model 1, the breakout time is around $2.55\ \rm d$, which is shorter by a factor of $\sim 17\%$. Generally, as seen in Fig. \ref{fig:velocity_models}, the shock velocity obtained by simulations is somewhere in between the piecewise model and the \citetalias{MM99} model. We compare to the numerical results in sec. \ref{sec:numerical}.

\textbf{Radiation-dominated density profile:}
For the piecewise velocity in the radiation-dominated profile (Eq. \ref{eq:rho_rad}),  the integrals are similar. They have the form 
\begin{equation}
    J(y,\alpha)=\intop y^{3/2}\left ( 1-\frac{y^2}{6} \right)^{3\alpha}dy=\frac{2}{5}y^{2/5} {}_{2}F_{1}\left(\frac{5}{4},-3\alpha,\frac{9}{4}, \frac{y^2}{6}\right),
\end{equation}
where ${}_2F_1(a,b,c,x)$ is the hypergeometric function. The breakout time is 
\begin{align}
    \left. \frac{t_{\rm bo}}{H/v_0} \right|_{\rm pw} = & \zeta_1 + \frac{\mathcal{A}_1}{\zeta_1^{3/2}} \left[  J(\zeta_2,3\mu') - J(\zeta_1,3\mu') \right] \nonumber \\
     + & \frac{\mathcal{A}_2}{\zeta_1^{3/2}} \left[  J(\zeta_{\rm bo},3\mu) - J(\zeta_2,3\mu) \right].
\end{align}

For the \citetalias{MM99} velocity model in the radiation-dominated profile, the breakout time is given by 
\begin{equation}
    \left. t_{\rm bo} \right|_{\rm MM99}= \frac{H}{v_{0}}\intop_{0}^{\zeta_{{\rm bo}}}\left(1+\frac{\rho_{0}H^{3}}{M_{{\rm ej}}}\zeta^{3}\right)^{1/2}\left(1-\frac{\zeta^{2}}{6}\right)^{3\mu}d\zeta,
\end{equation}
which is, again, not analytic but can be estimated numerically.

\textbf{Off-plane explosions:} For off-plane explosions the integrals are the same but the integration domain is $(\zeta_0, \zeta_{\rm bo})$, where $\zeta_0$ is the explosion site above the midplane. For the \citetalias{MM99} model this is the only change. 

For the piecewise model, we divide the domain into three regions, corresponding to the piecewise velocity solution in Eq. \ref{vpi}, namely 
\begin{equation}
    t_{\rm bo} = \frac{H}{v_0}(I_1+I_2+I_3).
\end{equation}
Here, $I_1=\zeta_1-\zeta_0$, is the integral over the (constant velocity $v_0$) free expansion phase;  $I_2$ is the integral between $\zeta_1$ and $\zeta_2$, where the velocity scales with the density with a power law $v\propto \rho^{-\mu'}$, where $\mu'=\mu +1/5$; and $I_3$ is the integral between $\zeta_2$ and $\zeta_{\rm bo}$, where the velocity scales with the density with a power law $v\propto \rho^{-\mu}$. 

For $\zeta_{\rm bo}> \zeta_2>\zeta_1>\zeta_0$, the result is essentially the same as in Eq. \ref{t_pw}, but with $I_1$ replacing $\zeta_1$ in the first term. This is the case for model 9.

For $\zeta_1< \zeta_{\rm bo}<\zeta_2$, the shock breaks out before the Sakurai acceleration begins. This leads to $I_3=0$ and the domain of integration of $I_2$ is $(\zeta_1, \zeta_{\rm bo})$. This is the case for models 10 and 11.

For $\zeta_{\rm bo} < \zeta_1$, the shock breaks out before the free expansion phase ends. In this case, we only have $I_1= \zeta_{\rm bo} - \zeta_1$. This is the case for model 12.
\section{CSM mass} \label{app:int_mass}

Here we explicitly calculate the integral in Eq. (\ref{eq:m_int}) for various models. We will start with  mid-plane explosions and hence fix $z_{\rm min}=0$.

For a flat profile, $z_{\rm max}=H$ and the CSM mass is simply $M_{\rm CSM}=4 \pi \rho_{0}H^{3}/3$.

For the Gaussian profile we have 

\begin{align}
M_{\rm CSM}(z_{\rm max}) & =4\pi\rho_{0}\intop_{0}^{z_{\rm max}}e^{-z^{2}/2H^{2}}z^{2}dz \nonumber \\
 & =4\pi\rho_{0}H^{3}\left[\sqrt{\frac{\pi}{2}}{\rm erf}\left(\zeta_{\rm max}/\sqrt{2}\right)-\zeta_{\rm max}e^{-\zeta_{\rm max}^{2}/2}\right] \label{eq:int_mass_gaussian}
\end{align}
where $\zeta_{\rm max}=z_{\rm max}/H.$ Note that for $\zeta_{\rm max}\to\infty$ we
get $M_{\rm{CSM},\infty}=(2\pi)^{3/2}\rho_{0}H^{3}$ while for small
$\zeta_{\rm max}\ll1$ it is essentially as the uniform density case, $M_{\rm CSM}(z_{\rm max}\ll H)\approx4\pi \rho_{0}z_{\rm max}^{3}/3$.
For all practical purposes, the value at the photosphere is close enough to the limiting value.

For the radiation-dominated profile, we have 
\begin{align}
M_{\rm CSM}(z_{\rm max}) & =4\pi\rho_{0}\intop_{0}^{z_{\rm max}}\left(1-\frac{z^{2}}{6H^{2}}\right)^3z^{2}dz \nonumber \\
 & =4\pi\rho_{0}H^{3}\left[\frac{\zeta_{\rm max}^{3}}{3}-\frac{\zeta_{\rm max}^{5}}{10}+\frac{\zeta_{\rm max}^{7}}{84}-\frac{\zeta^{9}_{\rm max}}{1944}\right] \label{eq:int_mass_rad}
\end{align}
For the maximal value $\zeta_{\rm max}=\sqrt{6}$ we have $M(z_{\rm max})=4\pi\rho_{0}H^{3}\times\frac{32}{35}\sqrt{\frac{2}{3}}=9.38\rho_0 H^{3}$. 

\textbf{Off plane explosions:} 
For an off plane explosion which occurs at $z_{\rm \rm min}$, the density ranges from $\rho(z_{\rm min})$ to $\rho(z_{\rm {max}} - z_{\rm min})$, which the vertical scale increases from $0$ to $z_{\rm max} - z_{\rm min}$. The resulting integral is
\begin{equation}
    M_{\rm CSM}(z_{\rm min}, z_{\rm max}) = 4\pi \intop_{0}^{z_{\rm max} - z_{\rm min}} \rho(z+z_{\rm min}) z^2 dz.
\end{equation}

For the Gaussian profile we have
\begin{align}
    \frac{M_{\rm CSM}}{4\pi \rho_0 H^3} = & \sqrt{\frac{\pi}{2}}(1+\zeta_{{\rm min}}^{2})\left[{\rm erf}\left(\frac{\zeta_{{\rm max}}}{\sqrt{2}}\right)-{\rm erf}\left(\frac{\zeta_{{\rm min}}}{\sqrt{2}}\right)\right] \nonumber \\
     - & e^{-\zeta_{{\rm min}}^{2}/2}\zeta_{{\rm min}}-e^{-\zeta_{{\rm max}}^{2}/2}(\zeta_{{\rm max}}-2\zeta_{{\rm min}})
\end{align}
where $\zeta_{\rm max/min} = z_{\rm max/min}/H$.
For model 9 we use $\zeta_{\rm max} = \infty$ and $\zeta_{\rm min}=1$, and the resulting mass is $M_{\rm CSM} = 1.92 M_\odot$. For model 10 we use $\zeta_{\rm max} = \infty$ and $\zeta_{\rm min}=2$, and the resulting mass is $M_{\rm CSM} = 0.15 M_\odot$.

For the radiation dominated profile we have 
\begin{align}
    \frac{M_{\rm CSM}}{4\pi \rho_0 H^3} = & \intop_{\zeta_{\rm min}}^{\sqrt{6}}(\zeta-\zeta_{{\rm min}})^{2}\left(1-\frac{\zeta^{2}}{6}\right)^{3}d\zeta = \frac{\zeta_{\rm min}^9}{54432} - \frac{\zeta_{\rm min}^7}{1260} \nonumber \\ + &  \frac{\zeta_{\rm min}^5}{60} - \frac{\zeta_{\rm min}^3}{3}+\frac{16\sqrt{6}\zeta_{\rm min}^2}{35}  -\frac{3\zeta_{\rm min}}{2} + \frac{32\sqrt{2}}{35\sqrt{3}}. 
\end{align}
For model 11 we use $\zeta_{\rm min}=1$, and the resulting mass is $M_{\rm CSM} = 0.49 M_\odot$. For model 12 we use $\zeta_{\rm min}=2$, and the resulting mass is $M_{\rm CSM} = 6.4\cdot10^{-4} M_\odot$.


\section{lightcurve computation in the 2D models} \label{app:lightcurve_hormone}

We first define an observer plane placed above the disc, parallel to the disc plane. From behind the disc, we solve the simple radiative transfer equations along parallel lines of sight to obtain the intensity at each position on the observer plane by computing the sequence
\begin{equation}
 I(z+\Delta z)=I(z)e^{-\Delta\tau}+S(1-e^{-\Delta\tau}),
\end{equation}
where $I$ is the intensity, $S$ is the source function and $\Delta\tau=\rho\kappa\Delta z$ is the optical depth along a small distance $\Delta z$. We assume black body radiation for the source function ($S=\sigma T^4/\pi$) and only compute the bolometric luminosity. For opacity, we only assume electron scattering ($\kappa=0.2(1+X)$~cm$^2$~g$^{-1}$, where $X$ is the hydrogen fraction). We do not take into account the effect of recombination in order to avoid making the disc unreasonably optically thin due to the way we set the initial disc temperature. This choice should not affect the early evolution of the lightcurve, at least in the planar phase. The intensity is then integrated over the whole observer plane and multiplied by $4\pi$ to obtain the luminosity. We refer the reader to \citet[]{suzuki2016} for more details of the idea behind the calculation procedure. 

Because radiation transport is not coupled to the hydrodynamics in our 2D simulations, this methodology does not account for the diffusion and radiative cooling in the ejecta.
 Therefore, the lightcurves obtained this way may not be energetically self-consistent. Additionally, the temperature in the outer parts of the ejecta may be overestimated. If the outer parts of the ejecta have sufficiently cooled, the atoms may recombine and lower the opacity, affecting the later lightcurve. Nevertheless, this estimate should give us a rough idea of how the shock breakout may be observed.

\bsp	
\label{lastpage}
\end{document}